\documentclass[11pt]{article}
\usepackage{fullpage}

\usepackage[utf8]{inputenc}
\usepackage{amsmath}
\usepackage{graphicx}

\usepackage{multirow}
\usepackage{rotating}

\usepackage{url}
\usepackage{mdframed}
\usepackage{xspace}
\usepackage{lipsum}
\usepackage{graphicx}
\usepackage{caption}
\usepackage{authblk}
\usepackage{url}

\usepackage{hyperref}

\usepackage{todonotes}

\usepackage{booktabs} 

\newcommand{\Paragraph}[1]{\noindent{\em #1}}

\begin{document}



  \title{Classifying Eyes-Free Mobile Authentication Techniques\thanks{A version of this paper is to appear in the Journal of Information Security and Applications (JISA)}}

  \author[*]{Flynn Wolf}
  \author[$\S$]{Adam J. Aviv}
  \author[*]{Ravi Kuber}
  \affil[*]{University of Maryland, Baltimore County}
  \affil[$\S$]{United States Naval Academy}
  \affil[ ]{\small {\tt \{flynn.wolf,rkuber\}@umbc.edu}, {\tt aviv@usna.edu}}
  


\maketitle
  
  \begin{abstract}

Mobile device users avoiding observational attacks and coping with situational impairments may employ techniques for eyes-free mobile unlock authentication, where a user enters his/her passcode without looking at the device. This study supplies an initial description of user accuracy in performing this authentication behavior with PIN and pattern passcodes, with varying lengths and visual characteristics. Additionally, we inquire if tactile-only feedback can provide assistive spatialization, finding that orientation cues prior to unlocking do not help. Measurements of edit distance and dynamic time warping accuracy were collected, using a within-group, randomized study of 26 participants. 1,021 passcode entry gestures were collected and classified, identifying six user strategies for using the pre-entry tactile feedback, and ten codes for types of events and errors that occurred during entry. We found that users who focused on orienting themselves to position the first digit of the passcode using the tactile feedback performed better in the task. These results could be applied to better define eyes-free behavior in further research, and to design better and more secure methods for eyes-free authentication.

\end{abstract}




\section{Introduction}
The threat of observational attacks in shared or public spaces may influence or
modify the way smartphone users interact with their devices.  In particular,
users may favor unlocking their mobile devices out-of-view, without looking at
the screen to avoid others from {\em surfing} the authenticator. Purposeful user
obfuscation (e.g. keeping the screen out of sight from third parties or hidden
cameras by hiding the device in the pocket or bag
\cite{abdolrahmani2016empirical}) for purposes of the initial stages of the
interaction, limits the likelihood of the authentication sequence being
viewed. This can put users at some level of ease, even if the remainder of the
interaction is performed in-view of third parties.

{\em Eyes-free authentication} behaviors may also be performed when the
situation, context or environment demands it. For example, in situations where
glare may be factor, or the environment is inappropriate for mobile device usage
and discretion is needed (e.g. \cite{grinter2006chatting}, the interaction may
be performed away from view).
While eyes-free interactions for different types of mobile device have been
studied by researchers in the past
\cite{azenkot2013digitaps,Azenkot:2013:EUS:2513383.2513440, bonner2010no,
  Brewster:2003:MEI:642611.642694, Goel:2012:WUA:2207676.2208662,
  gupta2011svift, li2008blindsight,
  kajastila2013eyes,Lu:2017:BET:3120957.3090083,pielot2012pocketmenu,Zhao:2007:EEM:1240624.1240836},
studies have yet to examine real world eyes-free authentication behaviors;
investigating the performance with common authentication mechanisms when the
phone is out-of-view, and user coping strategies to enter passcodes in an
eyes-free manner.

To address this knowledge gap, we conducted a randomized, multi-factor study
with 26 participants entering PINs and gesture-based patterns (termed: "patterns"
in this paper). Participants entered passcodes under both in-view and eyes-free
conditions, as well as eyes-free using an additional training module for
spatialization based on tactile feedback.

The tactile channel was chosen to discreetly offer cues directly to the user's
hand, without drawing attention during interaction, as would likely occur with
auditory or visual cues. Existing assistive aids aid to eyes-free PIN
authentication, such as iOS VoiceOver, rely on audio feedback (audio readout of
PIN number buttons when touched, allowing selection). However, audio cues impose
usability and security penalties in shared and public spaces.

Biometric authentication such as fingerprint identification can greatly expedite
this task for many users. However, fingerprint identification remains only a
secondary means of authentication, which is generally tied to a PIN or patterns for
screen unlocking. Essentially, even biometric authentication users must
necessarily enter conventional passcodes on a semi-regular basis, and eyes-free
conditions may apply in some instances.

In light of this, tactile-only feedback was designed for this study as a
research device for understanding authentication performance with strictly
eyes-free interaction. Its functionality, and our evaluation of its performance,
is not intended to propose a workable real-world tool in the present
form. Instead, we tried to capture how users develop techniques that use
additional spatial cues to locate key screen features. This spatialization might
then assist the accuracy and precision of eyes-free authentication gestures,
especially for situations where the user may feel at risk of being a victim of
an observer attack or be at risk of a situational impairment.

Given these assumptions, we have undertaken these research questions:
\begin{itemize}

\item {\bf RQ1}: How well are users able to perform eyes-free authentication (without tactile feedback) with common methods, such as PIN and pattern entry, and how is this affected by the length and visual features of passcodes?

\item {\bf RQ2}: Will the relationship between spatial cues to screen layout features (e.g. position of buttons), presented by tactile interaction, enhance  the user's performance when authenticating eyes-free?

\item {\bf RQ3}: When tactile feedback is presented, what approaches will users develop for using it?
\end{itemize}

With these considerations, during the experiment we collected complete movement traces, recording all participants' touch-based gestures during each authentication attempt, totaling 1021 eyes-free traces.  
To extend the work described in \cite{wolfetal2018}, we aimed to understand the input techniques and strategies the participants developed when completing the tasks.  To do this, we classified all the traces, and developed a set of verified and grounded labels to describe the actions of the participants. 

We further evaluated participants' performance in the eyes-free setting in two dimensions, accuracy and precision. For accuracy, we considered the edit-distance (or {\em Levenshtein distance}) between the input passcode and the true passcode. The edit-distance considers the number of additions or removals to transform one string sequence into another. For precision, we developed  a geometric distance measure between in-view and eyes-free traces using {\em Dynamic Time Warping} (DTW), computing the average distance between temporally-associated points in the trace. 

Based on this analysis, we found that participants using patterns were more accurate and precise in eyes-free settings, as compared to PINs. Additional tactile training was found to not improve the accuracy or precision of the participants' entries. We discuss users' observations regarding this distinction between task performances.
When applying the classification results, we found that specific techniques in both training stages impacted performance. In particular, traces where participants used the additional tactile training aid to understand specifically the location of the starting digit of their passcode showed the most significant increase in performance, for both PINs and patterns. In addition to identifying techniques that improved performance, we also developed a set of classifications for eyes-free entry and training. 

The results firstly contribute to an initial baseline of performance results and classifications of types for eyes-free interaction behaviors, events, and error types. We also show that the describe strategies for locating the starting location of authentication gestures (i.e. the screen position of the button for a passcode's first digit) that correspond with a number of significant effects on user performance. These results will help further research on eyes-free interaction make accurate comparisons and descriptions regarding this condition. Additionally, these insights will help iterate the design of targeted training aids for users, such as blind mobile technology users who rely on secure ubiquitous computing for privacy-sensitive tasks in shared spaces, who need to authenticate frequently in eyes-free settings (i.e. when at perceived risk of an observer attack described in \cite{abdolrahmani2016empirical}). Informing users of effective techniques will enable users to enter unlock authentication more confidently, securely, and accurately, away from adversarial observation.

While the tactile aid adopted for this study produced a mostly negative result
from accuracy and edit distance measures, we assert several important
contributions from this investigation:

\begin{enumerate}
  \item A novel characterization of HCI and security performance conditions for eyes-free authentication tasks.  
 \item A systematic inquiry of accuracy, precision, and timing effects of input in eyes-free settings. 
 \item Establishing the unequivocal performance gap between eyes-free PIN and pattern entry (although unsurprising, this is the first time this has been shown empirically).
 \item The extension of existing classification methodology for coding eyes-free unlock entry methods and events, similar to error codes established in von Zezschwitz et al. \cite{von2013patterns}.  
 \item	Identifying significant relationships between classification codes and authentication conditions (e.g. a decrease in Start-Hunt behavior for pattern passcodes ($\chi^2=8.17, p<0.005$)).
 \item	Identifying passcodes features for which accuracy and/or precision significantly deviated from average (e.g. self-crossing pattern 743521).
\end{enumerate}

We feel the relationship between the initial training methods that users develop
using the tactile aid, such as those that help locate the starting point of the
authentication gesture, are particularly illustrative. Strategies, such as the
Start-Hunt trial code and Return to Start training code, offer an insight into
the ways that users cope with the challenges of entering gestures under
eyes-free conditions.  By being able to better understand user strategies taken,
along with events and error types made, this work could lead to the improved
support of targeted training aids for users who interact with mobile
authentication solutions under eyes-free conditions.



\vspace{-.1in}
\section{Related Work}
\subsection{Eyes-Free Interaction Techniques}
As mobile technologies reduce in size and provide increasing amounts of PC-like functionality, these technologies become an attractive option for performing tasks while on-the-go.  As information is predominantly presented via the graphical user interface, the user is heavily reliant on visual feedback to perform mobile tasks.

However, there are scenarios when difficulties are faced viewing the interface.  One of the predominant issues relates to worries about third parties viewing content, and using this information without permission.  Examples described by Yi et al. \cite{Yi:2012:EUM:2207676.2208678}  include (1) environmental factors (e.g. excessive brightness impacting the user's ability to perceive screen content, and in scenarios where switching visual attention between the device and the physical environment poses safety concerns), (2) social factors (e.g. instances where it may be socially-inappropriate to view the screen, or multi-task in front of others), (3) constraints imposed by the mobile devices themselves (e.g. difficulties seeing content due to the crowded nature of content on mobile GUIs), and (4) personal factors (e.g. no perceived benefit to using vision to performing the task).

Additionally, if the user feels under threat of observer attacks, the screen may be hidden from view, either shielded by the hand \cite{ketabdar2012magnetic}, or placed within a garment or accessory \cite{abdolrahmani2016empirical}.  The user can then attempt to use a combination of a mental image of the interface and muscle memory to attempt to interact with the device.  

One of the fundamental motivations for eyes-free interaction is that as it leaves visual attention unoccupied, users are free to perform additional tasks \cite{oakley2007designing}. However, performing mobile tasks when visual and other forms of feedback are not available, can lead to errors during the input process, and contribute to levels of frustration among users. To better support users, a range of techniques have been developed involving gestural input (e.g. \cite{azenkot2013digitaps, bonner2010no,Brewster:2003:MEI:642611.642694, Goel:2012:WUA:2207676.2208662,Lu:2017:BET:3120957.3090083, Tinwala:2010:ETE:1868914.1868972, Zhao:2007:EEM:1240624.1240836}), or voice input (e.g. \cite{Azenkot:2013:EUS:2513383.2513440,li2008blindsight}), along with accessible forms of output to provide feedback to the user (e.g. audio \cite{bonner2010no,Brewster:2003:MEI:642611.642694, kajastila2013eyes, Tinwala:2010:ETE:1868914.1868972, Zhao:2007:EEM:1240624.1240836}, and/or tactile output \cite{gupta2011svift, Tinwala:2010:ETE:1868914.1868972} either to the user's hand via the mobile device or via a separate wearable.  Similar technologies have also been designed to support users for whom the visual channel is restricted or blocked (i.e., individuals who are blind and visually-impaired \cite{azenkot2012passchords,   guerreiro2012exploring, guerreiro2015blind, kane2013touchplates, mcgookin2008investigating}, with the aim of substituting or complementing visual feedback with other forms of accessible information).  However, any solutions developed would ideally need to work in conjunction with existing assistive technologies (e.g., screen readers).

\subsection{Augmenting Interfaces to Support Eyes-Free Interactions}
To help users orient position and better understand the layout of content on a mobile interface during eyes-free interactions, non-visual cues have been added to the interface. The PocketMenu solution \cite{pielot2012pocketmenu} for pocket-based mobile device interactions, presents tactile information  to convey the position and state of buttons, arrayed along the left edge of the mobile phone touchscreen, to exploit the natural tactile localization that the outer bezel of the phone affords. Findings from the researchers' study showed that PocketMenu outperformed auditory output (VoiceOver) in terms of completion time, selection errors, and subjective usability, making it ideal for interactions where the user is on-the-go \cite{pielot2012pocketmenu}.  McGookin et al. \cite{McGookin:2008:ITA:1463160.1463193} used tactile overlays to support interaction with a visual touchscreen. Recommendations proposed to support eyes-free interactions include presenting a discernible button for orientation (i.e. home button), through the use of a tactile aid such as an adhesive bump-on, and providing feedback for all interactions.  However, care should be taken with the latter, to ensure that the user is not overloaded with feedback.  Oakley and Park \cite{oakley2007designing} note the trade-off which designers should be aware of - between the amount of information contained within user interface feedback, the speed with which this can be achieved and the amount of effort and attention required to interpret it.  This is thought to be especially important in the eyes-free domain. Existing research has not directly addressed tactile feedback using built-in actuators for eyes-free authentication, that would orient users to any passcode starting position. 
 
\vspace{-.1in}
\subsection{Device and interaction modification for authentication}
Although not designed specifically to improve accessibility of entry under eyes-free conditions, researchers have adopted a series of methods to reduce the likelihood of adversarial observations. Examples include the  Back of Device scheme, proposed by De Luca et al. \cite{de2012touch}, who used a simulated double-sided touchscreen to allow users to grip the device in two hands, and conceal a secondary authentication action. This was found to be more resistant to simulated observation attacks than other common login methods. In terms of presenting information through alternative senses, the tactile channel has been a popular method of communicating information relating to the authentication discreetly to the user.  Examples include the Haptic Wheel \cite{Bianchi:2010:HWD:1753846.1754029}, where the user positions their hand around a rotary dial.  After each input, the system randomizes the vibration it emits to protect the user from observer attack, and the VibraPass system \cite{de2009vibrapass}, where the system presents tactile cues to the user's hands to indicate when to enter false PIN numbers to randomize each entry. While this modification added little time to a typical interaction (averaging 1.68 additional seconds), randomized PINs were intercepted by the simulated shoulder surfers 32.5\% of the time (versus 100\% for non-VibraPass patterns).  Other tactile solutions include H4Plock, proposed by Ali et al. \cite{ali2016developing}, where the user is required to enter a sequence of up to four pre-selected on-screen  gestures  while responding to tactile prompts signaling whether stimuli from a primary or secondary passcode should be entered. The solution proved to be secure against 76.5\% of participants, who carried out attacks immediately after watching a set of videos. Participants were able to express strong levels of confidence in using the system. 

Limited work has been conducted exploring existing common mobile authentication mechanisms and their use when the user needs to purposely obfuscate the screen to perform tasks in an eyes-free manner.  In this paper, we describe a study investigating PIN and graphical pattern entry using tactile feedback, when the device is out of view, with the aim to unlocking entry. While auditory feedback appears to be an appealing solution to this scenario, it may be impractical. Providing an auditory representation of screen content may be insecure or an unacceptable distraction.  Furthermore, auditory cues from a mobile device may be masked by ambient noise, which may pose challenges when attempting to authenticate entry. Tactile feedback may offer a solution to directing users to make accurate entry gestures (e.g. finding the start position of their passcode). A tactile aid to support orientation has also been evaluated, as part of this research.  To better understand user behaviors, we aim to focus on input techniques, and methods of classifying these.


\begin{figure*}[t]
  \centering
  \includegraphics[width=\linewidth]{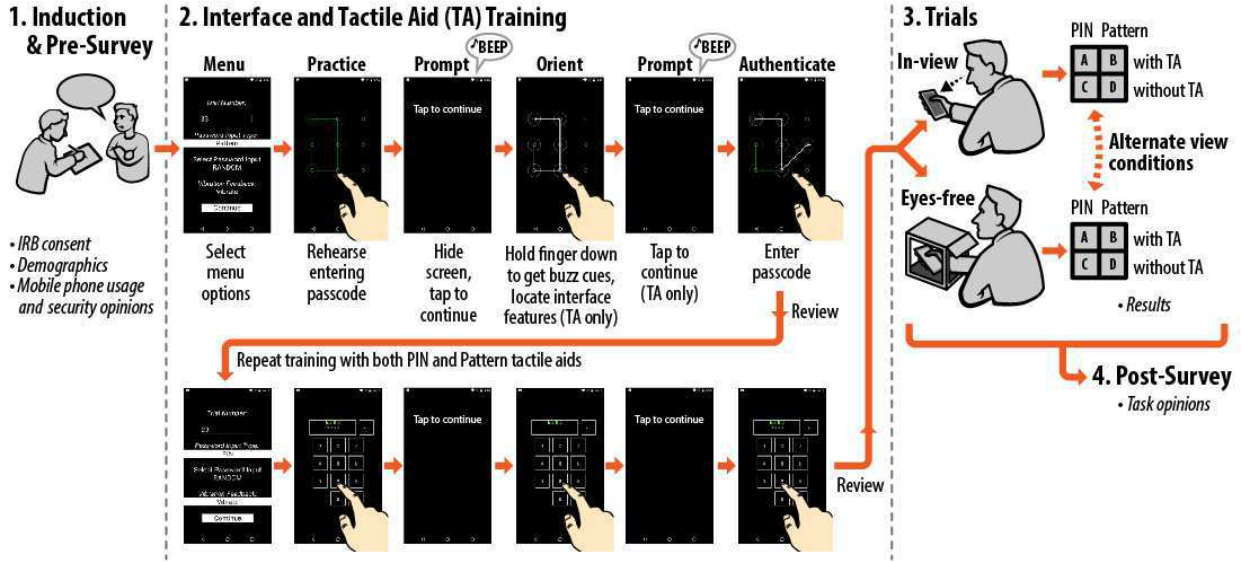}
  \caption{Study procedure: Participants are informed of the make-up of the experiment, and
    basic demographic and usage data is collected in step (1). In step (2),
    participants are trained on the application interface for data collection
    for both PINs and patterns, and spend time interacting with the training aid (the
    first prompt/orient phase). Without the training aid, participants in trial
    A and B proceed directly to the next phase. After completing the training stage, step (3)
    the trials begin alternating between the conditions as presented in Table~\ref{tab:conditions}.}
  \label{fig:procedure}
\end{figure*}

\begin{table}[t]
\centering
\small
\resizebox{\linewidth}{!}{%
\begin{tabular}{ c | c | c | c }

  \multicolumn{2}{c|}{Eyes-Free} & \multicolumn{2}{|c}{In-View}\\
\hline
\hline 
  \multicolumn{2}{c|}{A) w/o tactile aid} & \multicolumn{2}{|c}{B) w/o tactile aid}\\
\hline
   A1) 4-digit PIN & A3) 4-len Pattern &  B1) 4-digit PIN & B3) 4-len Pattern \\
   A2) 6-digit PIN & A4) 6-len Pattern &  B2) 6-digit PIN & B4) 6-len Pattern \\
  \multicolumn{2}{c|}{} & \multicolumn{2}{|c}{}\\
  \multicolumn{2}{c|}{C) w/ tactile aid} & \multicolumn{2}{|c}{D) w/ tactile aid}\\
\hline
  C1) 4-digit PIN & C3) 4-len Pattern &  D1) 4-digit PIN & D3) 4-len Pattern \\
  C2) 6-digit PIN & C4) 6-len Pattern &  D2) 6-digit PIN & D4) 6-len Pattern \\
\end{tabular}}
\caption{Conditions for study}
\vspace{-.1in}
\label{tab:conditions}
\end{table}

\section{Study Design and Procedure}

We designed a within-subjects, multi-factor study where participants were asked to enter authentication sequences under four primary conditions. To address RQ1, out-of-view performance was measured and compared to that of in-view. RQ2 is addressed by the inclusion of with- and without-tactile feedback conditions. The conditions are described in more detail below:
\begin{itemize}
\itemsep0em 
\item {\em in-view w/o training aid}: participants performed the tasks completely in-view where they could view the device and the entry thereon, without a training aid.

\item {\em in-view w/ training aid}: participants performed the tasks completely in-view with a tactile/vibration-based training aid.

\item {\em eyes-free w/o training aid}: participants performed the
  tasks out-of-view (eyes-free) without a training aid.
  
\item {\em eyes-free w/ training aid}: participants performed the task
  with a tactile/vibration-based training aid.
\end{itemize}

Within each condition, the participants were assigned a sequence of 10 PINs or patterns to enter, of which 5 were length 4 PINs/patterns and 5 were length 6 PINs/patterns. We summarize these conditions and the set of authentication passcodes in Table~\ref{tab:conditions} and Table~\ref{tab:pass}, respectively. Passcodes were sourced from real world data to establish validity, and selected individually to include important visual characteristics for analysis, for example left vs. right side shift, and self crossing patterns. For all experiments, we used a Nexus 5x, which has a common 5.79" x 2.86” form factor, 5.2” display, 1080x1920 resolution. The steps of the study are illustrated in Figure~\ref{fig:procedure}. The procedure and study design are similar to that presented in related work ~\cite{tinwala2008letterscroll, von2015easy}. 

\subsection{Study Procedure} 
As presented in Figure~\ref{fig:procedure}, the study consists of three steps: (1) introduction/pre-study, (2) training stage, and (3) experimental trials. In total, the procedure took about 40 minutes per participant, as training time was allocated to familiarize participants with relevant authentication issues, the two training aids, and the methods of the study. This was undertaken to prepare participants at least up to a low level of performance that could simulate everyday usage.
To begin, in step (1), participants provided informed consent of participation (as required by the IRB), and were asked a series of pre-survey questions. This included demographic questions as well as questions regarding their mobile phone usage and locking habits. 

The training phase followed. As we developed a specialized data collection application that directed and stepped users through the procedure (see Figure 1 for more details), it was necessary to provide training and familiarization with the interface to allow participants to focus on accuracy and precision using the tactile feedback rather than any unnecessary memorization tasks. In particular, we wished for participants to be fully aware of each of the authentication codes (i.e., all the PIN and pattern examples that they would encounter so they would be familiar with them when entering them in an eyes-free manner). A visual reference of the passcode to be entered was always available throughout all conditions, via a laptop showing the passcode. We also wished to train participants on the tactile aid that some of the trials would employ (condition C and D in Table~\ref{tab:conditions}), so that they could maximize the use of the aid. We also wanted to give participants some simulated trial runs under eyes-free conditions, where they could subsequently review their performance by viewing replays of their entries, provided within the data collection application. 
Once participants were comfortable with the interface, the procedure, and the tactile feedback aid, the trials for each condition began. The conditions were randomized using a Latin Square. 

\subsection{Interface and Tactile Feedback Design}
A web-based application was developed using HTML5 to collect data for the study. When opened in a smartphone's web browser, the application interfaces closely simulated the layout, font, animation, luminance, color, and interaction of typical mobile authentication screens. The pattern interface displayed the standard Android OS 3x3 grid, and the PIN interface displayed the standard 0-9 digit layout (telephone dial-pad style), including a text display that showed entered digits (which are obfuscated to an ``*'' after 1 second), a back/delete button next to the display, and an ``OK'' button to the right of the ``0'' to submit the PIN. Graphical depictions of the application are presented in Figure~\ref{fig:procedure} with the study's procedures. 

The data collection application’s interface is designed to assist the study procedure by presenting the participant with an authentication code and then directing users to place the device in the eyes-free setting to complete the experiment. Additionally, to support RQ3, the interface implemented the tactile feedback aid if the condition called for that. The interface application also collected and stored traces of all touch interactions. A trace consisted of the x- and y-coordinate of a touch event and the time in which that event occurred. To direct participants in the eyes-free setup, the interface played audible ``beeps'' to indicate to the participants transition between the steps of the experiment. 

\begin{figure}
\centering
  \includegraphics[width=0.5\columnwidth]{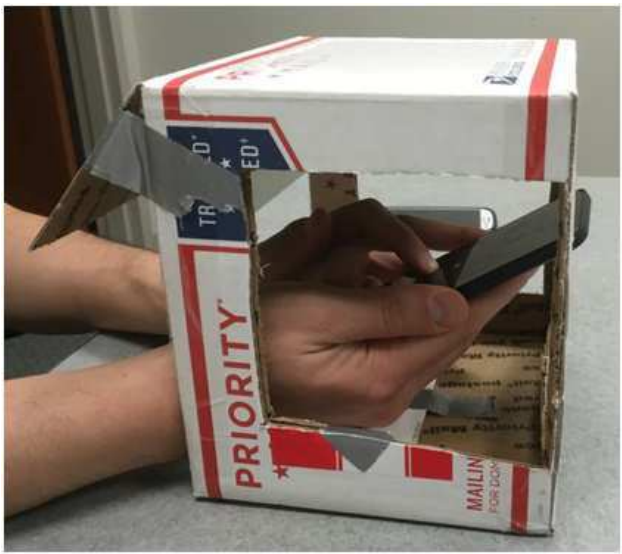}
  \caption{Participant holding the LG Nexus 5x test phone inside the eyes-free
    observation box.}
  \label{fig:shield}
\end{figure}

To facilitate the eyes-free setting, we used a shielding box in which the participant can interact with and hold the device in a natural posture, but cannot see the device itself (see Figure~\ref{fig:shield}). Cut-outs in the sides of the shielding box allowed the researchers to observe the interaction, which the participant is unable to view. We placed no restrictions on how the user chose to interact with the device, i.e, using one or two hands, as long as they were not able to view the device directly. 

The interface also incorporated tactile feedback in the form of vibrations for touches that occurred within a digit/point. This feature was not originally included in our experiment, and was added after prototyping the procedures based on participant feedback. We believe this is a reasonable choice given that most unlock authentication systems already incorporate this style of feedback.

In the conditions with the tactile training aid, participants were able to use the aid to orient with the interface prior to beginning the eyes-free authentication stage. The orientation occurred during continual touching, where a participant can swipe around the screen, receiving tactile/vibrational feedback when swiping over a digit/point. If the digit/point is the first digit/point in the PIN/pattern, a vibration cue encoded with a faster rhythm was used to differentiate this point from the others; however, no other passcode specific feedback was provided. Once the participant lifts their finger, ending the continuous swiping, the interface “beeps” indicating they should attempt to enter the authentication. Trace information was collected during tactile training phase as well in the same format as in the authentication phase.

\subsection{Pre- and Post-Survey Procedures}
In steps (1) and (3), we asked participants a set of pre- and post-survey questions. Copies of the questions are  presented in the Appendix. The questions included basic demographic information, as well as inquiries regarding mobile device usage, such as OS, time, and locking behavior. Reported statistics are available in Table~\ref{tab:demo}. 

In the pre-survey portion, participants were also asked a series of questions about how they chose passcodes, their concern for phone security, experiences with device theft or unauthorized access, and how those factors might have affected their behavior. Two questions were Likert responses, on a scale from 1-5: 
\begin{itemize}
\itemsep0em 
\item {\em How concerned are you with keeping your phone secure (1, not at all
  concerned, to 5, highly concerned)?}
\item {\em How concerned are you typically, in public spaces, with the threat of
  someone watching you authenticate and collecting your passcodes (1, not at all
  concerned, to 5, highly concerned).}
\end{itemize}

Responses to these questions did not significantly correlate with performance on the task, but did provide some information on the participants usage and locking behavior and mindset that is relevant to the task. The responses are also presented in Table~\ref{tab:demo}. 

Following the tasks, during the post-survey, we asked questions regarding the ease or difficulty of the task itself, with or without the tactile feedback aid. These results are presented in the results section of the paper.

\begin{table}[t]
\centering
\small
\begin{tabular}{ c| r || r c }
  & PINs   & Patterns & \multirow{11}{*}{\hspace{+.2in}\fbox{\includegraphics[width=0.25\linewidth]{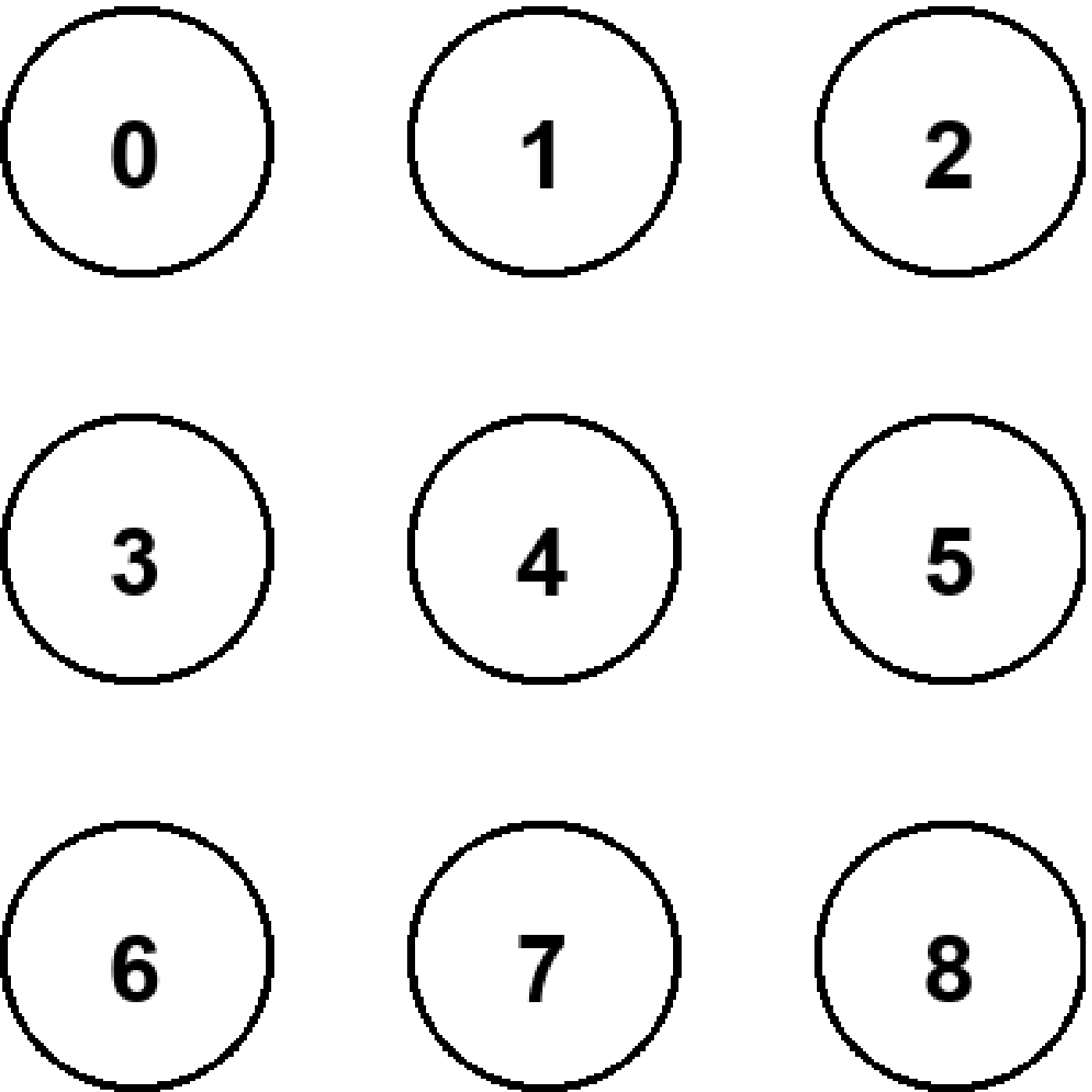}}}\\
  \cline{1-3}
  \multirow{5}{*}{4-length} & 1328   & 0145 \\
  & 1955   & 1346 \\
  & 5962   & 3157 \\
  & 6702   & 4572 \\
  & 7272   & 6745 \\
  \cline{1-3}
  \multirow{5}{*}{6-length} & 153525 & 014763 \\
  & 159428 & 136785 \\
  & 366792 & 642580 \\
  & 441791 & 743521 \\
  & 458090 & 841257 \\
  \end{tabular}
  \caption{PINs and patterns used in experiments, and reference for labeling pattern contact points. See the Appendix and \url{https://www.dropbox.com/s/m1ujrtt32a7blg2/suplm.zip?dl=0}{Supplementary Materials} for graphical depictions of the authentication}.
\label{tab:pass}
\end{table}

\subsection{Passcode Selection}
To select the passcodes, both PINs and patterns, we sought to use a representative sample from real world data for validity, that also had good spatial features in terms of the regions of the screen that must be touched. As such, we used the set of PINs and patterns published in related work~\cite{aviv2017baseline,Davin:2017:BMS:3027063.3053221}. The set of passcodes are presented in Table~\ref{tab:pass} and visuals are provided in the Appendix and the \url{https://www.dropbox.com/s/m1ujrtt32a7blg2/suplm.zip?dl=0}{Supplementary Materials \footnote{Supplementary materials - https://www.dropbox.com/s/m1ujrtt32a7blg2/suplm.zip?dl=0}}.

Following~\cite{aviv2017baseline}, the PINs from the study were culled from publicly leaked password sets, using sequences of 4 or 6 digits. This is an acceptable practice in simulating PINs from related work~\cite{Bonneau2012}, as good realistic data is not available for purposes of research. Patterns from~\cite{aviv2017baseline} were selected from~\cite{aviv2015isbigger} using spatial criteria, where the authors ensured good spatial criteria: up, down, left, right, and neutral shifts. 

\begin{table}[t]
  \centering
  \small
  \begin{tabular}{ c r | r | r | r}
    & & Male & Female & Total \\
    \hline
    & total & 12 & 14 & 26\\
    \hline
    \multirow{2}{*}{\rotatebox{90}{age}}& 18-24 & 7 & 7 & 14\\
    & 24-34 & 5 & 7 & 12\\
    \hline
    \multirow{2}{*}{\rotatebox{90}{OS}} & Android & 8 & 5 & 13\\
    & iOs & 4 & 9 & 13\\
    \hline
    \multirow{4}{*}{\rotatebox{90}{\parbox{1.7cm}{\centering Unlock \\Choice}}} & Fingerprint & 5 & 10 & 15\\
    & PIN-6 & 2 & 7 & 9 \\
    & PIN-4 & 5 & 4 & 9\\
    & Pattern & 4 & 2 & 9\\
    & No-Lock & 1 & 1 & 2\\
    \hline
    \multirow{4}{*}{\rotatebox{90}{\parbox{2cm}{\centering Level of Confidence\\(1-5 Likert)}}} & Phone Security & $3.75$& $3.79$ & $3.77$ \\
    &                & (STD: 1.42) &(STD: 0.7) & (STD: 1.07) \\
    & Shoulder Surfing & $3.08$ & $3.79$ &  $3.46$ \\
    &                  & (STD: 1.24) & (STD: 0.97) & (STD: 1.140\\
  \end{tabular}
\smallskip
\caption{Demographics of participants}
\label{tab:demo}
\end{table}

\subsection{Participants and Recruitment}
Participants were recruited via a university mailing list, alongside the use of the snowball sampling technique.  The latter was used with the aim of recruiting a wider, more representative group of mobile device users.  There were 26 individuals recruited, comprising of 12 male and 14 female younger adults.  Participants were fully sighted mobile users, evenly divided between using iOS and Android. For their current unlock conditions, many used their fingerprint in addition to another lock mechanism, such as a PIN or pattern. Only 2 participants had no lock authentication on their smartphone. Details are provided in Table 3. 

\subsection{Realism and Limitations}
As the study was performed under controlled in-person conditions, the set-up introduced some constraints which reduced the level of realism. Examples include the slightly extended posture of the user’s arm, holding and pointing with the mobile device when interacting within the shielding box. Further work will examine more natural methods of interacting with the device under more realistic constraints (e.g. the impact of the user entering passcodes while the device is located in his/her pocket or bag). 

In order to provide a baseline control, participants were asked to perform in-view PIN and pattern conditions, with/without the presence of a tactile aid. Due to the randomization of conditions, half of the group of participants performed this condition after the eyes-free condition. This was conducted to minimize the likelihood of an order effect. It was acknowledged that this may have led to a slight performance disadvantage for users performing eyes-free first, as the in-view conditions may be considered equivalent to a small amount of extra training. 

In terms of limitations, passcodes were presented in the same order to all participants under each condition. Although passcodes did not necessarily increase in complexity during each condition, it is acknowledged that this may have contributed to an effect.  


While the number of participants recruited for our study either exceeded or remained in line with those selected for other studies relating to eyes-free interaction (e.g. \cite{pielot2012pocketmenu,McGookin:2008:ITA:1463160.1463193}), the sample selected for the study described in this paper, can be considered small in size (n=26), and limited in range of demographic features. While significance was found within our results, challenges can be faced when samples are very small, which can impact the power of the study and increase the margin of error.  Similarly, the representativeness of findings may be impacted if the samples are too limited in terms of age.  The sample in our study reflected the university environment (i.e. younger adults). It is also acknowledged that participants who were primarily iOS users may have been at a slight disadvantage in levels of experience interacting with stroke-based patterns, as these types of mechanism are found on devices running the Android OS. However, training was provided to familiarize users with the process, which helped to offset any potential disparity. Finally,as participants came from a university, issues of bias may have also been introduced from the working environment.  While it is not uncommon in authentication studies to recruit participants comprising mainly of students, future studies would aim to widen the pool, with the aim of recruiting a more diverse sample, who are more representative of the array of individuals who use mobile devices (e.g., participants with varying levels of concern about security).


\section{Results}
In this section, we present the results of both the performance of the participants, as well as the classification of the traces and training techniques. We begin with a description of the metrics applied, followed by the performance results, and finish with the classification results. 

\subsection{Performance Metrics}

\Paragraph{Accuracy.} 
A crucial and informative metric to determine effectiveness is simply how
accurately participants performed the tasks in eyes-free settings. Of course,
authentication is a binary response: either participants entered the passcodes
correctly or not. As we are also interested in granularity of performance, we also
considered the {\em edit distance}, normalized to the distance of each
passcode. More precisely, we considered accuracy as a fraction calculated for a
passcode $p$ and entered code $p'$
\begin{equation*}
  acc = \frac{len(p) - d(p,p')}{len(p)}
\end{equation*}
where $len(p)$ is the length of the passcode and $d(p,p')$ is the edit distance
between the entered and expected passcode. The edit distance (or Levenshtein
distance) computes the number of additions, subtractions, or replacements needed
to transform one sequence into another. For example, if the task requires
entering the passcode 123456, and the participant entered 12356 (or any passcode
off by one in some dimension), then the accuracy would be $(6-1)/6$ or $0.83$ as
the edit distance is 1. This is a generous accuracy measure in the sense that
the edit distance is a greedy algorithm and tries to aggressively match
strings. However, given the nature of the task, eyes-free entry, we feel that
this provides a better reflection on participant effort and performance than a
binary yes/no.

\begin{figure}
\centering
  \begin{tabular}{c c}
    {\small \bf (a)} & {\small \bf (b)}\\
    \fbox{\includegraphics[width=0.35\linewidth]{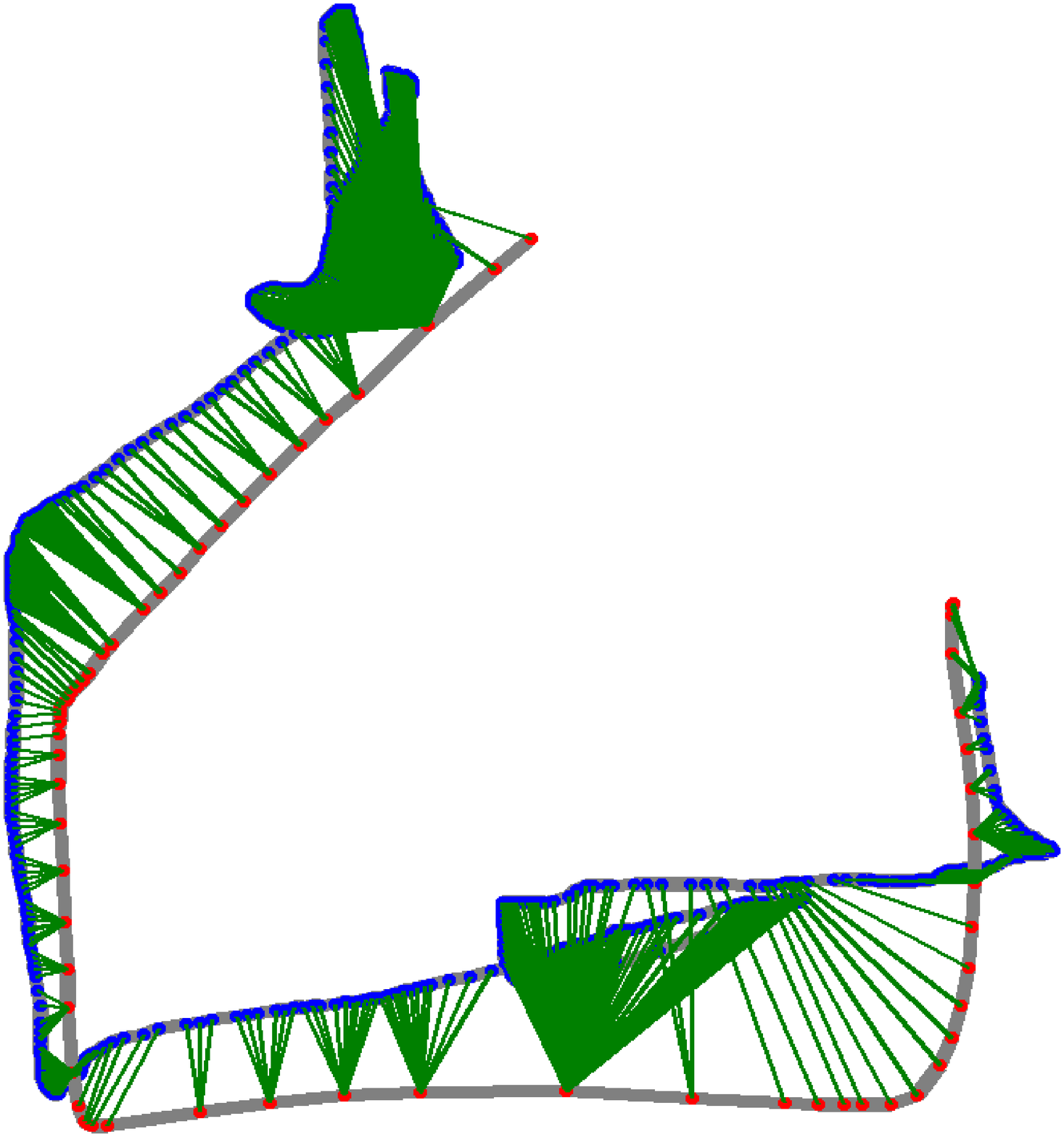}} &
                                                                               
\fbox{\includegraphics[width=0.35\linewidth]{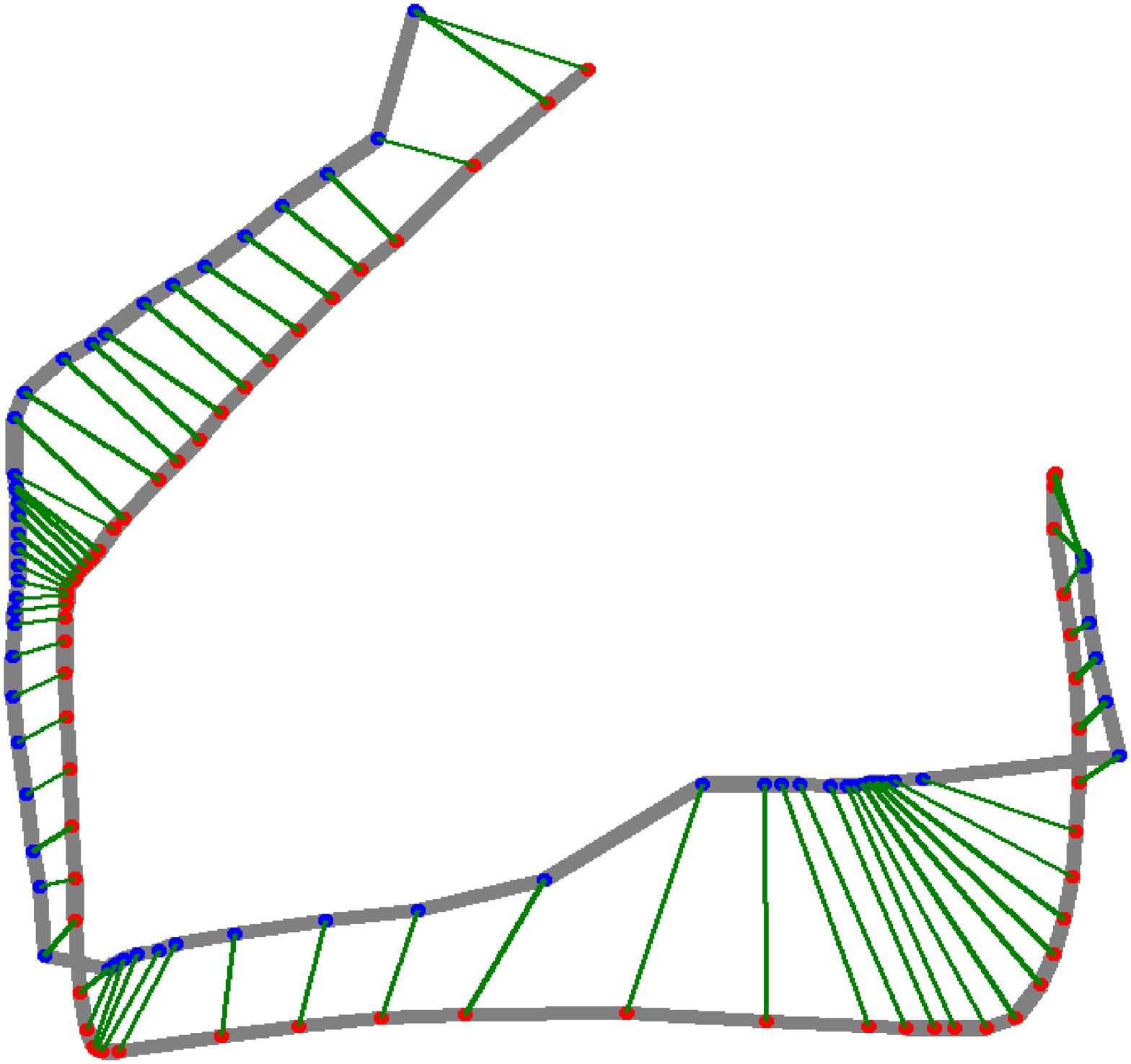}} \\
  \end{tabular}
  \caption{Example of using DTW to do point reduction: In (a) The blue trace is
    eyes-free and the red trace is in-view, and the green lines show the points
    in the blue trace time sequence matched to points in the red trace; and in
    (b), associated points in the blue trace are merged via Euclidean average to
    form a set of singly matched points to the red trace. The precision metric
    is then the average Euclidean distance between the matched point of the blue
    trace with the red one. }
  \label{fig:dtw}
\end{figure}

\begin{sidewaystable}[p]
  \centering
\small

\begin{tabular}{ c | c | c | c | c}
  & Tactile Aid & Pattern & PIN  & $t$-test\\
\hline
\multirow{2}{*}{\em acc.} & w/o & ($\mu=0.80$, $\sigma=0.28$, $n=250$) & ($\mu=0.72$, $\sigma=0.30$, $n=260$) & $t=3.08$,$p<0.05*$ \\
 & w/ & ($\mu=0.83$, $\sigma=0.27$, $n=253$) & ($\mu=0.71$, $\sigma=0.30$, $n=258$) & $t=4.78$,$p<0.001**$ \\
 & $t$-test  &  $t=1.17$,$p=0.243$  &  $t=-0.49$,$p=0.628$  & \\
\hline
\multirow{2}{*}{\em prec.} & w/o & ($\mu=80.93$, $\sigma=60.40$, $n=250$) & ($\mu=145.18$, $\sigma=90.29$, $n=260$) & $t=-9.41$,$p<0.001**$ \\
 & w/ & ($\mu=83.21$, $\sigma=69.74$, $n=253$) & ($\mu=154.49$, $\sigma=94.11$, $n=258$) & $t=-9.71$,$p<0.001**$ \\
 & $t$-test  &  $t=0.39$,$p=0.695$  &  $t=1.15$,$p=0.251$  & \\
\hline
\multirow{2}{*}{\em start} & w/o & ($\mu=80.51$, $\sigma=67.62$, $n=250$) & ($\mu=96.56$, $\sigma=109.79$, $n=260$) & $t=-1.98$,$p<0.05*$ \\
 & w/ & ($\mu=68.35$, $\sigma=61.96$, $n=253$) & ($\mu=91.77$, $\sigma=86.46$, $n=258$) & $t=-3.51$,$p<0.001**$ \\
 & $t$-test  &  $t=-2.10$,$p<0.05*$  &  $t=-0.55$,$p=0.582$  & \\
\hline
\multirow{2}{*}{\em time (ms)} & w/o & ($\mu=7134.86$, $\sigma=6672.95$, $n=250$) & ($\mu=8597.35$, $\sigma=6314.62$, $n=260$) & $t=-2.54$,$p<0.05*$ \\
 & w/ & ($\mu=7500.12$, $\sigma=5359.63$, $n=253$) & ($\mu=10227.07$, $\sigma=8236.91$, $n=258$) & $t=-4.43$,$p<0.001**$ \\
 & $t$-test  &  $t=0.68$,$p=0.499$  &  $t=2.53$,$p<0.05*$  & \\
\end{tabular}
\caption{Performance results: {\em acc.} is the accuracy using the edit-distance measure, {\em prec.} is the precisions using the DTW method, {\tt start} is the Euclidian distance of the start point, and {\tt time} refers to the number of milliseconds. As the data followed a normal distribution, we used a two-tailed $t$-test. Horizontally, the $t$-test compared Pattern vs. PIN results, and vertically, the $t$-test compared w/ and w/o the tactile aid. Effect size of $\alpha=0.05$ was considered significant. Only traces that were complete and collected without errors were considered.}
\label{tab:perf}

\end{sidewaystable}

\begin{sidewaystable}[p]
\begin{tabular}{ c c | c | c | c}
Pattern & Metric & w/o Tactile Aid & w/ Tactile Aid & $t$-test \\
\hline  
\multirow{2}{*}{ 0145 } & {\em acc} & ($\mu=0.81$, $\sigma=0.27$, $n=25$) & ($\mu=0.86$, $\sigma=0.27$, $n=25$) & $t=-0.66,p=0.510$ \\
     & {\em prec} & ($\mu=82.66$, $\sigma=50.16$, $n=25$) & ($\mu=71.77$, $\sigma=62.11$, $n=25$) & $t=0.68,p=0.499$ \\
\hline
\multirow{2}{*}{ 1346 } & {\em acc} & ($\mu=0.88$, $\sigma=0.19$, $n=25$) & ($\mu=0.87$, $\sigma=0.22$, $n=26$) & $t=0.25,p=0.801$ \\
     & {\em prec} & ($\mu=57.19$, $\sigma=28.84$, $n=25$) & ($\mu=72.96$, $\sigma=37.22$, $n=26$) & $t=-1.69,p=0.098$ \\
\hline
\multirow{2}{*}{ 3157 } & {\em acc} & ($\mu=0.78$, $\sigma=0.31$, $n=25$) & ($\mu=0.75$, $\sigma=0.28$, $n=26$) & $t=0.36,p=0.717$ \\
     & {\em prec} & ($\mu=89.60$, $\sigma=77.38$, $n=25$) & ($\mu=87.02$, $\sigma=45.19$, $n=26$) & $t=0.15,p=0.885$ \\
\hline
\multirow{2}{*}{ 4572 } & {\em acc} & ($\mu=0.92$, $\sigma=0.14$, $n=25$) & ($\mu=0.80$, $\sigma=0.29$, $n=25$) & $t=1.86,p=0.068$ \\
     & {\em prec} & ($\mu=67.62$, $\sigma=31.71$, $n=25$) & ($\mu=130.00$, $\sigma=125.20$, $n=25$) & $t=-2.41,p=0.020$* \\
\hline
\multirow{2}{*}{ 6745 } & {\em acc} & ($\mu=0.68$, $\sigma=0.38$, $n=25$) & ($\mu=0.74$, $\sigma=0.42$, $n=25$) & $t=-0.53,p=0.595$ \\
     & {\em prec} & ($\mu=111.61$, $\sigma=96.17$, $n=25$) & ($\mu=90.51$, $\sigma=80.56$, $n=25$) & $t=0.84,p=0.405$ \\
\hline
\multirow{2}{*}{ 014763 } & {\em acc} & ($\mu=0.81$, $\sigma=0.26$, $n=25$) & ($\mu=0.87$, $\sigma=0.24$, $n=26$) & $t=-0.74,p=0.461$ \\
       & {\em prec} & ($\mu=82.28$, $\sigma=68.34$, $n=25$) & ($\mu=73.37$, $\sigma=66.77$, $n=26$) & $t=0.47,p=0.640$ \\
\hline
\multirow{2}{*}{ 136785 } & {\em acc} & ($\mu=0.80$, $\sigma=0.23$, $n=25$) & ($\mu=0.86$, $\sigma=0.22$, $n=25$) & $t=-0.93,p=0.356$ \\
       & {\em prec} & ($\mu=70.12$, $\sigma=37.28$, $n=25$) & ($\mu=80.78$, $\sigma=53.51$, $n=25$) & $t=-0.82,p=0.418$ \\
\hline
\multirow{2}{*}{ 642580 } & {\em acc} & ($\mu=0.78$, $\sigma=0.25$, $n=25$) & ($\mu=0.79$, $\sigma=0.23$, $n=25$) & $t=-0.20,p=0.844$ \\
       & {\em prec} & ($\mu=78.69$, $\sigma=40.08$, $n=25$) & ($\mu=109.83$, $\sigma=76.32$, $n=25$) & $t=-1.81,p=0.077$ \\
\hline
\multirow{2}{*}{ 743521 } & {\em acc} & ($\mu=0.77$, $\sigma=0.27$, $n=25$) & ($\mu=0.92$, $\sigma=0.17$, $n=26$) & $t=-2.34,p=0.024$* \\
       & {\em prec} & ($\mu=89.66$, $\sigma=69.72$, $n=25$) & ($\mu=54.70$, $\sigma=26.51$, $n=26$) & $t=2.38,p=0.021$* \\
\hline
\multirow{2}{*}{ 841257 } & {\em acc} & ($\mu=0.80$, $\sigma=0.31$, $n=25$) & ($\mu=0.86$, $\sigma=0.22$, $n=24$) & $t=-0.79,p=0.434$ \\
       & {\em prec} & ($\mu=79.91$, $\sigma=49.08$, $n=25$) & ($\mu=62.14$, $\sigma=32.16$, $n=24$) & $t=1.49,p=0.142$ \\
\end{tabular}
\caption{Performance Metrics per-Pattern: Comparisons were made with-out (w/o) and with (w/) the tactile aid, and a $t$-test is used as the data is normal. Note that not all participants provided valid traces, and invalid traces were excluded.}
\label{tab:perf:pin}
\end{sidewaystable}

\begin{sidewaystable}
\begin{tabular}{ c c | c | c | c}
PIN & Metric & w/o Tactile Aid & w/ Tactile Aid & $t$-test \\
\hline  
\multirow{2}{*}{ 1328 } & {\em acc} & ($\mu=0.73$, $\sigma=0.32$, $n=26$) & ($\mu=0.69$, $\sigma=0.33$, $n=26$) & $t=0.43,p=0.672$ \\
     & {\em prec} & ($\mu=110.42$, $\sigma=54.18$, $n=26$) & ($\mu=108.75$, $\sigma=41.15$, $n=26$) & $t=0.12,p=0.901$ \\
\hline
\multirow{2}{*}{ 1935 } & {\em acc} & ($\mu=0.73$, $\sigma=0.36$, $n=26$) & ($\mu=0.62$, $\sigma=0.36$, $n=26$) & $t=1.06,p=0.296$ \\
     & {\em prec} & ($\mu=149.54$, $\sigma=125.19$, $n=26$) & ($\mu=169.98$, $\sigma=104.29$, $n=26$) & $t=-0.64,p=0.525$ \\
\hline
\multirow{2}{*}{ 5962 } & {\em acc} & ($\mu=0.82$, $\sigma=0.20$, $n=26$) & ($\mu=0.72$, $\sigma=0.40$, $n=26$) & $t=1.09,p=0.280$ \\
     & {\em prec} & ($\mu=136.19$, $\sigma=81.36$, $n=26$) & ($\mu=166.62$, $\sigma=119.08$, $n=26$) & $t=-1.08,p=0.287$ \\
\hline
\multirow{2}{*}{ 6702 } & {\em acc} & ($\mu=0.65$, $\sigma=0.29$, $n=26$) & ($\mu=0.63$, $\sigma=0.28$, $n=26$) & $t=0.24,p=0.810$ \\
     & {\em prec} & ($\mu=176.51$, $\sigma=98.64$, $n=26$) & ($\mu=172.79$, $\sigma=97.10$, $n=26$) & $t=0.14,p=0.891$ \\
\hline
\multirow{2}{*}{ 7272 } & {\em acc} & ($\mu=0.81$, $\sigma=0.31$, $n=26$) & ($\mu=0.67$, $\sigma=0.31$, $n=26$) & $t=1.56,p=0.124$ \\
     & {\em prec} & ($\mu=174.64$, $\sigma=112.77$, $n=26$) & ($\mu=176.16$, $\sigma=100.48$, $n=26$) & $t=-0.05,p=0.959$ \\
\hline
\multirow{2}{*}{ 153525 } & {\em acc} & ($\mu=0.78$, $\sigma=0.28$, $n=26$) & ($\mu=0.82$, $\sigma=0.21$, $n=26$) & $t=-0.66,p=0.515$ \\
       & {\em prec} & ($\mu=115.73$, $\sigma=62.08$, $n=26$) & ($\mu=136.91$, $\sigma=78.55$, $n=26$) & $t=-1.08,p=0.286$ \\
\hline
\multirow{2}{*}{ 159428 } & {\em acc} & ($\mu=0.69$, $\sigma=0.26$, $n=26$) & ($\mu=0.72$, $\sigma=0.25$, $n=26$) & $t=-0.45,p=0.654$ \\
       & {\em prec} & ($\mu=125.48$, $\sigma=53.41$, $n=26$) & ($\mu=144.00$, $\sigma=92.45$, $n=26$) & $t=-0.88,p=0.380$ \\
\hline
\multirow{2}{*}{ 366792 } & {\em acc} & ($\mu=0.66$, $\sigma=0.32$, $n=26$) & ($\mu=0.71$, $\sigma=0.24$, $n=26$) & $t=-0.65,p=0.517$ \\
       & {\em prec} & ($\mu=178.64$, $\sigma=99.21$, $n=26$) & ($\mu=196.46$, $\sigma=107.88$, $n=26$) & $t=-0.62,p=0.538$ \\
\hline
\multirow{2}{*}{ 441791 } & {\em acc} & ($\mu=0.72$, $\sigma=0.27$, $n=26$) & ($\mu=0.81$, $\sigma=0.17$, $n=24$) & $t=-1.46,p=0.151$ \\
       & {\em prec} & ($\mu=173.36$, $\sigma=84.65$, $n=26$) & ($\mu=163.50$, $\sigma=73.30$, $n=24$) & $t=0.44,p=0.663$ \\
\hline
\multirow{2}{*}{ 458090 } & {\em acc} & ($\mu=0.65$, $\sigma=0.32$, $n=26$) & ($\mu=0.71$, $\sigma=0.28$, $n=26$) & $t=-0.62,p=0.540$ \\
       & {\em prec} & ($\mu=111.25$, $\sigma=51.50$, $n=26$) & ($\mu=110.44$, $\sigma=53.00$, $n=26$) & $t=0.06,p=0.956$ \\
\end{tabular}
\caption{Performance Metrics per-PIN: Comparisons were made with-out (w/o) and with (w/) the tactile aid, and a $t$-test is used as the data is normal. Note that not all participants provided valid traces, and invalid traces were excluded.}
\label{tab:perf:pattern}
\end{sidewaystable}


\Paragraph{Precision.} 
As a second measure of performance, we wish to compare the traces of the
eyes-free entry to that of the in-view entry. Recall that interface application
collects detailed trace information for each touch event in the form of
$d=(x,y,t)$ where $x$ is a width coordinate, $y$ is a height coordinate, and $t$
is a time indication from the start of the interaction. The goal is to develop a
method that allows two sequences, which may be time dilated with different
number of points, to be compared in the 2-D space.

The solution to this was the use of Dynamic Time Warping (DTW), which has been
used extensively in the space of free-form gesture
authentication~\cite{yang2016freeform, liu2017guessing, clark2015egineering,
  de2012touch}. DTW takes two time series sequences and aligns one to the other
to best match the time dilation present. For example, in
Figure~\ref{fig:dtw}(a), the blue trace is the eyes-free trace and the red trace
is the in-view trace. Notice that the blue trace (eyes-free) is longer in time
(will have more points in the trace) and also have more loops due to mistakes
during entry. Comparing these two traces directly, such as point by point, would
indicate that the participant performed the authentication task quite poorly, but in fact, here, this is an
100\% accurate entry of pattern 136785.

With DTW, we can better show the precision of the user. First, DTW will associate
points in one time series with the other based on a distance measure, which is
Euclidean in this case. In Figure~\ref{fig:dtw}(a), this matching is represented
by the green lines. Based on the matching, point reduction is performed by
averaging the points in the blue trace that match a single point in the red
trace, giving us Figure~\ref{fig:dtw}(b). The last step in the metric is to take
the average distance between two matched points, again using Euclidian, which
represents the precision of the trace in the eyes-free setting to the in-view
one. 

We treated PIN traces as a series of distinct touch event traces since each
touch could really be a drag/swipe and consist of multiple points in the trace.
To apply the DTW distance metric for precision, we merged the touch event traces
into a single trace, via concatenation, and then applied the same routine as
described previously. This resulted in extra connections when the participant
transitioned between touch events, but this information is useful when
considering the precision as the direction and regularity of those transitions
are measurable and meaningful. We excluded the last touch event from the
analysis, pressing ``OK'' as this was not always successful in the eyes-free
conditions, and the researcher monitoring the experiment would need to assist in
this process.

\begin{sidewaysfigure}
\centering
\small
\begin{tabular}{ c c c c c c}
\fbox{\includegraphics[width=0.12\linewidth]{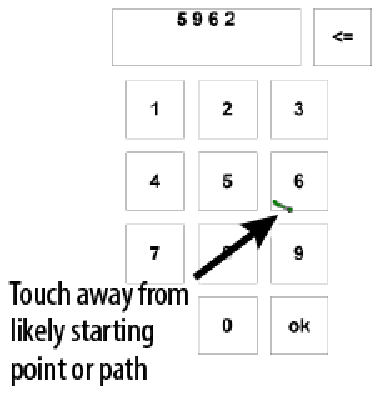}} &
\fbox{\includegraphics[width=0.12\linewidth]{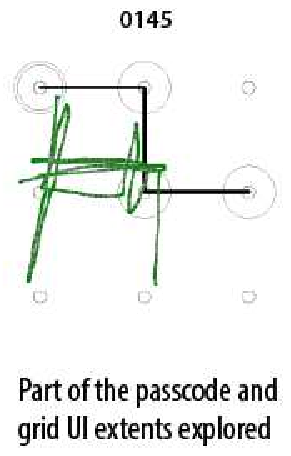}} &
\fbox{\includegraphics[width=0.12\linewidth]{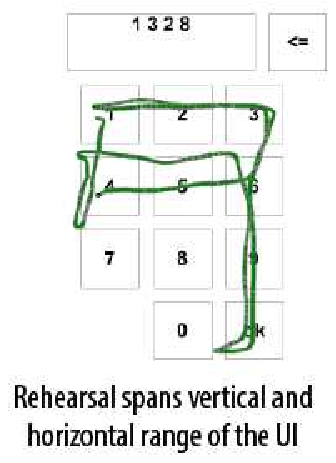}}&
\fbox{\includegraphics[width=0.12\linewidth]{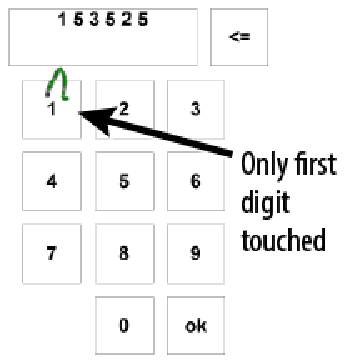}} &
\fbox{\includegraphics[width=0.12\linewidth]{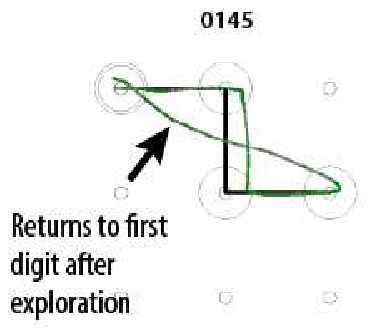}} &
\fbox{\includegraphics[width=0.12\linewidth]{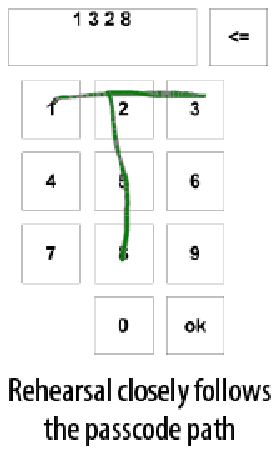}}\\
Random Touch & Partial Explore & Wide Explore & Start Point Only & Return to Start & Traced Path \\
\end{tabular}
\caption{Samples of Training Classification: see Table~\ref{tab:class:entry} for description of each classification type}
\label{fig:class:train}

\bigskip

\begin{tabular}{ c c c c c }
\fbox{\includegraphics[width=0.12\linewidth]{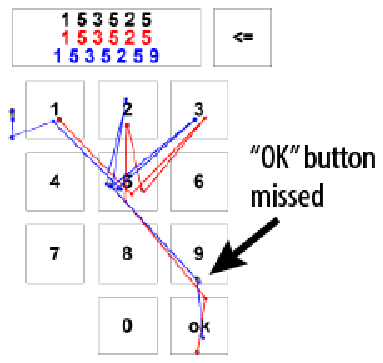}} &
\fbox{\includegraphics[width=0.12\linewidth]{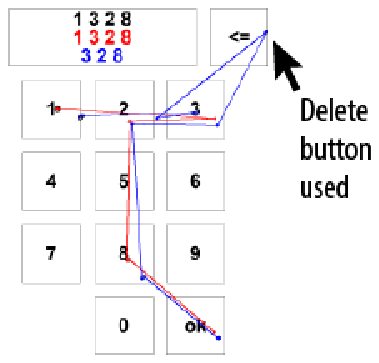}} &
\fbox{\includegraphics[width=0.12\linewidth]{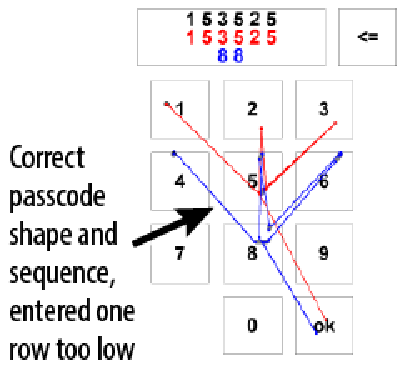}} &
\fbox{\includegraphics[width=0.12\linewidth]{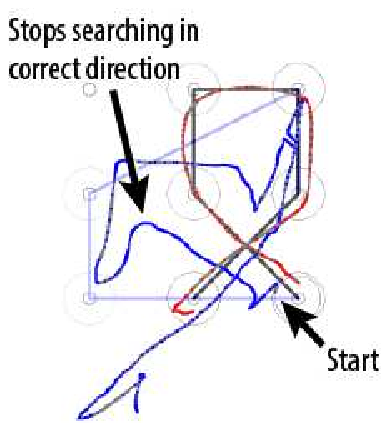}} &
\fbox{\includegraphics[width=0.12\linewidth]{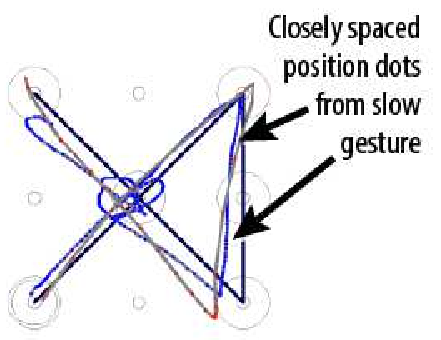}} \\
Missed OK (PIN) & Used Delete (PIN) & Transposed & Gave Up& Hesitant\\
\fbox{\includegraphics[width=0.12\linewidth]{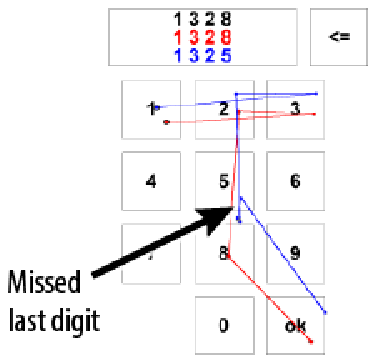}} &
\fbox{\includegraphics[width=0.12\linewidth]{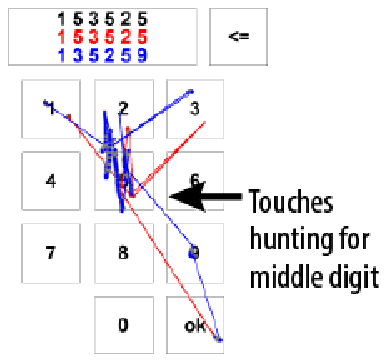}} &
\fbox{\includegraphics[width=0.12\linewidth]{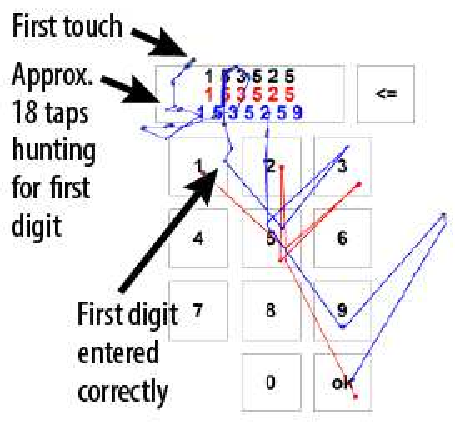}} &
\fbox{\includegraphics[width=0.12\linewidth]{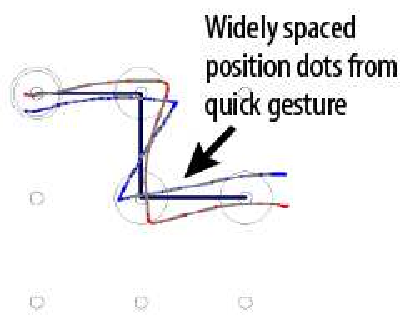}} &
\fbox{\includegraphics[width=0.12\linewidth]{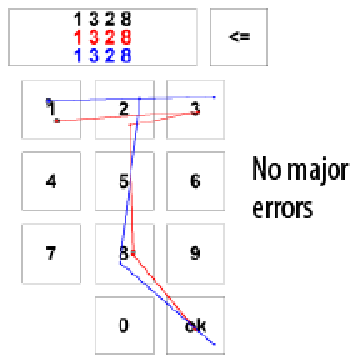}} \\
End-Hunt& Middle-Hunt& Start-Hunt& Swift & Went-Well\\
\end{tabular}
\caption{Samples of Entry Classification: see Table~\ref{tab:class:entry} for description of each classification type}
\label{fig:class:entry}
\end{sidewaysfigure}


\subsection{Performance Results}
The primary performance results are presented in Table~\ref{tab:perf} with per
passcode breakdowns in Table~\ref{tab:perf:pin} and
Table~\ref{tab:perf:pattern}. We applied the two metrics (accuracy and
precision) described previously to conditions that considered the use or non-use
of the tactile aid with the two unlock authentications, PINs and
patterns. Additionally, we considered two further measurements, the precision of
just the start point, calculated as the Euclidean distance between the first
touch event in the in-view to the eyes-free setting, and the time (in
milliseconds) of entering the authentication in the eyes-free setting.

As observed in Table~\ref{tab:perf}, the impact of the tactile aid is rather
limited. There were small effects for patterns, as compared to no effect in PINs,
and the effect was most notable when it comes to the start point in the
pattern. Using the aid showed a significant improvement in starting accuracy as
compared to not using the tactile aid. This is reasonable given that the tactile
aid performed different vibration feedback for the first point/digit in the
passcode. As the $t$-test performed is two-tailed, one can consider the
significance for the accuracy measure is somewhat intriguing,
$p=0.243/2=0.1215$. While this effect is not-significant as a one-tailed result,
it is encouraging for tactile aids in the eyes-free setting, perhaps a better
design based on feedback from this study, could improve the accuracy of entry.

The performance of the tactile aid for PINs can only be explained as detrimental. The eyes-free task for PINs is already significantly harder, in all conditions, but the addition of the aid showed worse performance for precision with no effect on accuracy (maybe even hurting accuracy). The time increase of using the aid is also striking and much larger than the time increase for patterns. We can speculate as to the reason for this disparity from participants’ post hoc responses regarding eyes-free PIN entry. It may be that the task is already challenging enough (concentrating on a series of discrete gestures hitting PIN digits) that the cognitive burden of integrating the aid’s tactile feedback only decreased the participants’ abilities. This is an area of future investigation, suggesting that different kinds of tactile aids may be needed for different authentication systems in the eyes-free setting. 

When observing performance impacts of the tactile aid on a per-passcode basis,
see Tables~\ref{tab:perf:pin} and~\ref{tab:perf:pattern}, there is no noticeable
effect for either accuracy or precision. In only one case, for pattern 743521,
was there a significant difference gained from using the tactile aid for both
accuracy and precision. This pattern required doubling-back, 
(see the \url{https://www.dropbox.com/s/m1ujrtt32a7blg2/suplm.zip?dl=0}{Supplemental Material} for a graphical depiction)
and the tactile aid may have provided some
reference points for that process which improved performance. However, there were
some negative results associated with using the tactile aid. For pattern 4572, there was an effect
observed for precision, as the tactile aid {\em increased} the precision
score, which degrades the precision of the pattern (smaller distances are
better). This may have been the case that due to the compactness of this
pattern, the tactile aid encouraged participants to take more elaborate/longer
paths than they would have normally. In either case, this is further evidence
that introducing aids for the eyes-free setting needs further investigation.

\subsection{Classification Metrics}
As a further metric for understanding the techniques that participants applied to eyes-free authentication, we developed a set of classification labels over both the authentication task and the tactile training. These labels describe observed types of errors (e.g. Transposed or Missed OK) or actions (e.g. Went Well or Start Hunt) made during passcode entry by participants, or apparent strategies used to spatialize using the tactile aid (e.g. Wide Explorer or Start Point Only). For consistency in label development, we adopted a similar approach to that of Micinski et al. \cite{micinski2017user} and von Zezschwitz et al \cite{von2013patterns}. Three researchers independently classified sub-sets, then met periodically to resolve differences between categorization labels. Although the definitions of some labels were mutually exclusive (e.g. a gesture could not be both Went Well and also include a Start Middle or End Hunt event), multiple compatible labels were applied to completely describe each gesture. Once agreement was reached on labels, one researcher labeled all the data while the other labeled a 15\% random sample. Comparing the ratings labels, there was strong agreement between the two researchers using Cohen’s $kappa$ ($\kappa= .900$, $p < 0.0005$). Additionally, during this procedure the researchers marked traces as invalid any observed miss-touches or other errors, which were then removed from the data sets.
%
%

The procedure produced six training labels and ten authentication labels. Visual
examples of these labels are provided in Figures~\ref{fig:class:train}
and~\ref{fig:class:entry} and descriptions of the labels, with counts, are
presented in Tables~\ref{tab:class:train} and~\ref{tab:class:entry}.
Animated examples of the classifications are provided on the \url{https://unlockclassification.wordpress.com}{supplementary web site\footnote{Supplementary web site - https://unlockclassification.wordpress.com}}.

\begin{table*}[t]
\centering
\small
\resizebox{\linewidth}{!}{
\begin{tabular}{ c|c | l | l }
  {\em Classification Name} & {\em Description} & {\em PIN} & {\em Pattern} \\
  \hline
  Random Touch & Short gesture away from path  & 4 (1.61\%) & 0 (0.00\%) \\
  \hline
  Partial Explore & Touched part of space or path & 100 (40.32\%) &  57 (22.98\%) \\
  \hline
  Start Point Only & Only traced the area around the first digit & 59 (23.79\%) & 47 (18.95\%)\\
  \hline
  Return to Start & Returns to starting point at end of gesture & 37 (14.92\%) & 37 (14.92\%)\\
  \hline
  Wide Explore & Touched most of the space & 32 (12.90\%) &  23 (9.27\%)\\
  \hline
  Traced Path  & Traced the path of the entire passcode & 54 (21.77\%) & 119 (47.98\%)\\
\end{tabular}}
\caption{Descriptions of the Training Classifications and Frequencies: Frequencies are calculated based on the total number of traces that received the label. Note that the trace can receive multiple labels, for example, it is possible to trace the path and return to the start. There was a total of 496 training traces considered, 248 pattern traces and 248 PIN traces.}
\label{tab:class:train}
\end{table*}

\begin{sidewaystable}
\begin{tabular}{ c|c| l | l | l | l } 
                            &                   &  \multicolumn{2}{| c|}{\em PIN} &  \multicolumn{2}{ c}{\em Pattern} \\
  {\em Classification Name} & {\em Description} & {\em w/o Tactile Aid} & {\em w/ Tactile Aid}& {\em w/o Tactile Aid} & {\em w/ Tactile Aid}\\
\hline
Start-Hunt & Missed or hunted for 1st digit/point & 61 (12.13\%) & 49 (9.74\%) & 66 (13.47\%) & 37 (7.55\%)\\
\hline
End-Hunt & Missed or hunted for last digit/point & 77 (15.31\%) & 69 (13.72\%) & 64 (13.06\%) & 66 (13.47\%)\\
\hline
Middle-Hunt & Missed or hunted for midrange digits & 104 (20.68\%) & 116 (23.06\%) & 95 (19.39\%) & 103 (21.02\%) \\
\hline
Went Well & No major errors & 111 (22.07\%) & 94 (18.69\%) & 105 (21.43\%) & 117 (23.88\%) \\
\hline
Gave Up & Just gave up and moved randomly & 2 (0.40\%) & 1 (0.20\%) & 2 (0.41\%) & 6 (1.22\%)\\
\hline
Transposed & Entered right shape in wrong place & 11 (2.19\%) & 3 (0.60\%) & 12 (2.45\%) & 3 (0.61\%)\\
\hline
Swift & Moved swiftly & 1 (0.20\%) & 5 (0.99\%) & 1 (0.20\%) & 1 (0.20\%)\\
\hline
Hesitant & Moved hesitantly & 2 (0.40\%) & 1 (0.20\%) & 1 (0.20\%) & 0 (0.00\%) \\
\hline
Used Delete (PIN-only) & Used Delete button to back up & 11 (2.19\%) &  8 (1.59\%) & - & - \\
\hline
Missed OK (PIN-only) & Tried to tap OK but missed >2-3 times & 43 (8.55\%) & 43 (8.55\%) & - & -\\

\end{tabular}
\caption{Descriptions of the Entry Classifications: Frequencies are calculated based on the total number of traces that received that label.  Note that the trace can receive multiple labels, for example, it is possible to trace the path and return to the start. There was a total of 993 entry traces considered, 490 pattern traces and 503 PIN traces.}
\label{tab:class:entry}
\end{sidewaystable}


\vspace{-.1in}
\subsection{Classification Results}
\vspace{-.05in}
The frequency results for the different classification labels are presented in
Tables~\ref{tab:class:train} and~\ref{tab:class:entry}. The frequencies are
calculated by the total traces in the categories that were labeled as
valid. Some traces were excluded due to either collection errors or
misunderstandings on the part of the participant.

For the training classifications (see Table~\ref{tab:class:train}), the most
common method for entering PINs was to perform a partial exploration of the interface. This
kind of exploration would entail a basic orientation of the space which would
make the participant aware of the digit-buttons. The next two most common
training classifications for PINs are just exploring the start point (Start Point Only, orienting to where to begin the PIN entry), or tracing out the entire gesture for entering all of the digits in the PIN (Traced Path). 

For patterns, during training, unsurprisingly, the most common technique we
observed was traces that retraced the path of the pattern. This makes sense
given the swiping nature of the pattern. Partial exploration of the grid space
and exploring just the start-point only are also quite common, but only half as
common as retracing.

When considering the classifications during entry, the traces are divided into
those that use the tactile aid and which do not. There was no difference
statistically in the presence of a label (using $\chi^2$) except for a
significant decrease in Start-Hunt for patterns ($\chi^2=8.17, p<0.005$). Since
the tactile aid provided distinct feedback when the participant touched the
start point, a decrease in the amount of hunting for the start point is
expected. While there was a similar decrease for PIN entry using the aid, it was
not significant ($\chi^2=1.31, p=0.25$).

\begin{sidewaystable}[p]
\centering
\small
\begin{tabular}{c | c | c | c | c}
                      & \multicolumn{2}{|c|}{\em PIN} & \multicolumn{2}{c}{\em Pattern} \\
{\em Classification} & {\em acc.} & {\em prec} & {\em acc.} & {\em prec.} \\
\hline
Random Touch & ($\mu=0.69, \sigma=0.32$) & ($\mu=142.75, \sigma=140.53$) & - & - \\
\hline
Partial Explore  & ($\mu=0.73, \sigma=0.29$) & ($\mu=139.14, \sigma=78.03$)  & ($\mu=0.78, \sigma=0.27$) & ($\mu=84.08, \sigma=47.74$)\\
\hline
Start Point Only & ($\mu=0.78, \sigma=0.31$) & ($\mu=126.27, \sigma=77.20$) & ($\mu=0.88, \sigma=0.18$) & ($\mu=75.29, \sigma=57.68$)\\
\hline
Return to Start & ($\mu=0.76, \sigma=0.24$) & ($\mu=124.19, \sigma=59.80$) & ($\mu=0.91, \sigma=0.22$) & ($\mu=66.41, \sigma=53.93$) \\
\hline
Wide Explore & ($\mu=0.69, \sigma=0.31$) & ($\mu=194.93, \sigma=114.63$) & ($\mu=0.80, \sigma=0.38$) & ($\mu=100.48, \sigma=94.65$) \\
\hline
Traced Path & ($\mu=0.67, \sigma=0.25$) & ($\mu=169.40, \sigma=88.19$) & ($\mu=0.86, \sigma=0.26$) & ($\mu=75.51, \sigma=57.97$) \\
\hline
            & $H=8.268, p=0.14$ &  $H=17.74, p<0.005**$ & $H=10.34, p<0.05*$ & $H=11.70, p<0.05*$\\
\end{tabular}
\caption{Performance Metrics for Training Classifications: Use Kruskal's H-test to test for significance.}
\label{tab:class:perf}
\end{sidewaystable}

In Table~\ref{tab:class:perf}, we analyze the impact of training label on
performance. For PINs, the accuracy shows no significant differences across
labels. However, we do see a difference for precision. Notably, both Start Point
Only and Return to Start have much lower distances (higher
precision). Additionally, those that performed a wide exploration had the
highest distance (lower precision). This suggests that training with PINs should
encourage a tighter training regime, perhaps confined to the start digit.

The training classification for patterns had significant differences for both
accuracy and precision. Most notably, the Return to Start label showed much
higher accuracy (0.91) and the lowest in the distance metric (highest
precision). We also see strong performance gains for both Start Point Only
and Traced Path. These results suggest that when designing training aids for
patterns, just like PINs, focusing on the start point is important. However, for
patterns, encouraging a retrace of the path of the pattern can also be helpful.

\subsection{Discussion from Participants}


Quantitative findings suggested that ease of learning and ease of use with the tactile aid were both harder for PIN compared to patterns (Table~\ref{tab:ease}),  In a similar vein, post-hoc discussion with participants revealed that challenges in the PIN condition were more frequently reported (n=7, three preferred the PIN aid operation). Several reasons for this were mentioned, most commonly the difficulty of accurately making long jumps between PIN numbers (e.g. 7 to 3 or 1 to OK, n=6). Participants also found more closely spaced PIN buttons a challenge (n=3), as well as accurately hitting the OK button at the end of a PIN sequence (n=3). Several specific passcode features, such as PIN digits located along the edge opposite a thumb for one-handed users (n=3), and consecutive numbers (n=1) were also mentioned as challenging factors. 

More generally, participants reported the 6-digit PIN and grid conditions as being more difficult to perform (n=6), and felt the app itself might be hard to use discreetly in public (n=5). There was positive feedback on the utility of the distinct start point feature (n=4), which could guide the user to accurately begin their gesture. Regarding this feature, one participant stated that “it’s all about the start” in maintaining spatialization throughout authentication gestures. Several participants also suggested giving the OK and Delete buttons a different vibrotactile coding to enhance the aid. 

Two participants related feeling the aid worked best with a two-handed index pointer grip, but this challenged them because they typically gripped their phones one-handed, with a thumb pointer. Another participant, also typically a one- handed thumb grip user, noted that PIN jumps to central digits (e.g. 5 and 8) were much easier than jumps to digits on the outer edges (e.g. 1, 7, and 9).

\begin{table}
  \centering
  \small
  \begin{tabular} { c | c | c}
                     & \em PIN & \em Pattern \\
    \hline
    Ease of Learning & $\mu=2.41, \sigma=1.25$ & $\mu=1.89, \sigma=0.93$\\
    \hline
    Ease of Use &  $\mu=3.44,\sigma=1.01$ & $\mu=2.74, \sigma=1.13$\\
  \end{tabular}
  \caption{Likert Responses to Ease of Tactile Aid: Response range from 1-5 (1, very easy, to 5, very hard).}
  \label{tab:ease}
\end{table}  



\vspace{-.1in}
\section{Conclusion}
This study was conducted to examine interaction techniques developed by users when they entered different types of passcodes on a mobile touchscreen device under eyes-free conditions (e.g., when worrying about the threat of observer attacks, or when facing situational impairments). We also inquired if tactile-only spatial feedback would effectively assist users with this type of screen unlocking. We did succeed, firstly, in capturing a picture of eyes-free authentication behaviors for common passcode entry methods. Looking at accuracy and precision measures, in particular, we can say regarding RQ1 that eyes-free unlocking overall (without tactile feedback) is understandably very challenging. PIN authentication is harder to perform with accuracy and precision, likely because of the numerous jumps the pointing finger must make.  Looking at these measures for gestures that were aided by tactile feedback, as addressed by RQ2, we see a small positive effect on start point accuracy, for pattern unlocking. Otherwise, the tactile only feedback employed in this study is not helpful to eyes-free authentication in terms of accuracy, precision, or time taken to unlock.  Movement traces were classified to better understand the strategies undertaken for unlocking and using tactile feedback, per RQ3. We found that strategies that focused on locating the first digit of the passcode were helpful. Precision for PIN entry was improved by the Start Point Only and Return to Start strategies. Similarly, accuracy and precision of pattern entry was helped by use of the Return to Start strategy.  These classifications have been described, and can be used by other researchers for purposes of analyzing movement traces on mobile touchscreen technologies. These insights about user strategies, event and error types can also support the design of targeted authentication training aids for users who may frequently encounter similar eyes-free conditions.

\section{Acknowledgments}
We thank Chukwuemeka KC Marume for his help with the data gathering and analysis process.  This work is supported by the Office of Naval Research.

\section{References}

\bibliographystyle{plain}
\bibliography{refs}

 \appendix







\section{Patterns and PINs Visualized}
\label{sec:viz}

\subsection{Patterns}
\label{fig:patterns}
The double circle indicates a start point, single circles is a point included in
the pattern. Note that labeling of patterns begins in the upper left with 0,
incrementing across each row, ending in the lower right with 8. All visuals are also
provided in {\tt images/patterns} sub-directory.
\begin{center}
\begin{tabular}{c c c c c }\small
\fbox{\includegraphics[width=0.15\linewidth]{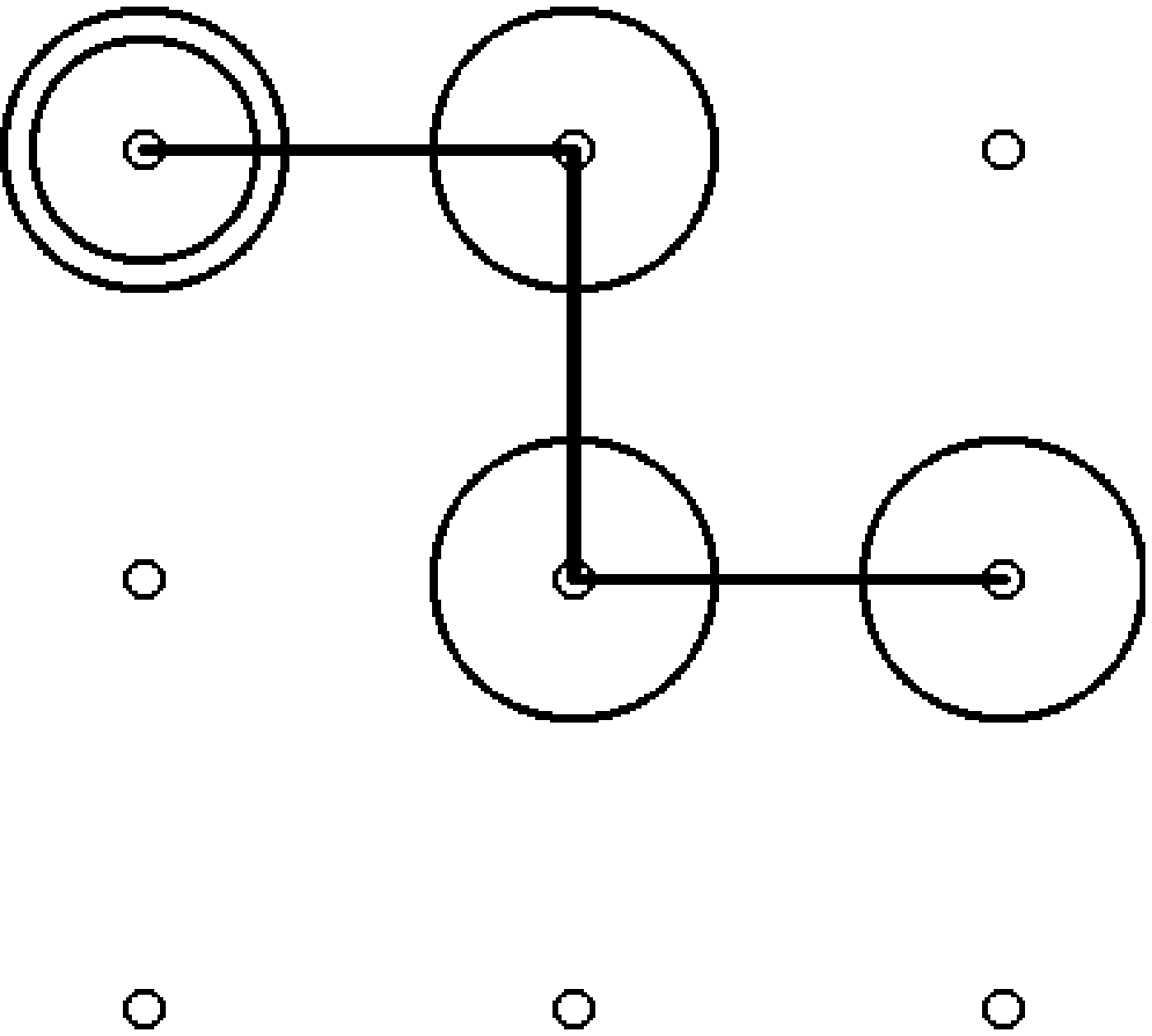}} &
\fbox{\includegraphics[width=0.15\linewidth]{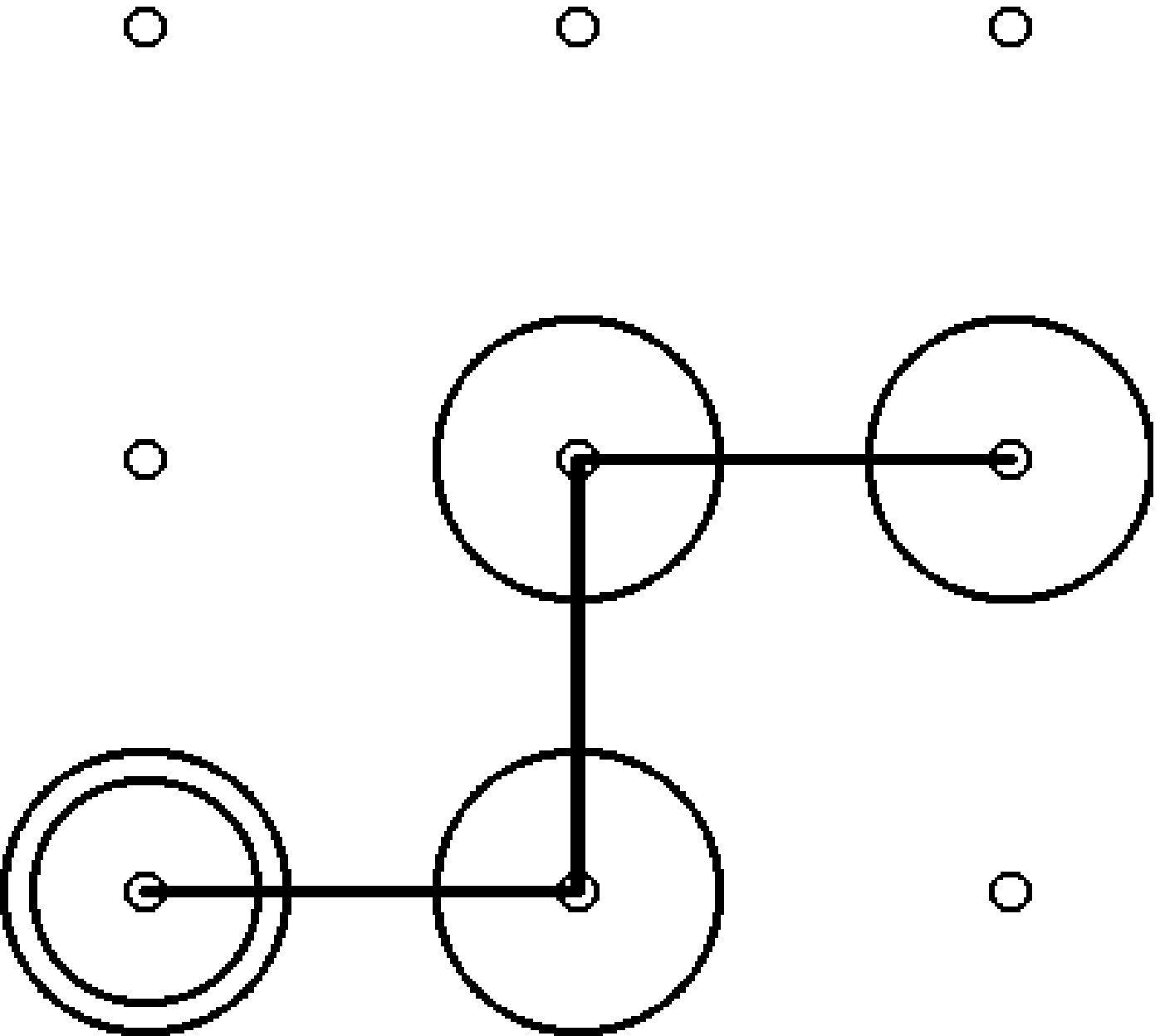}} &
\fbox{\includegraphics[width=0.15\linewidth]{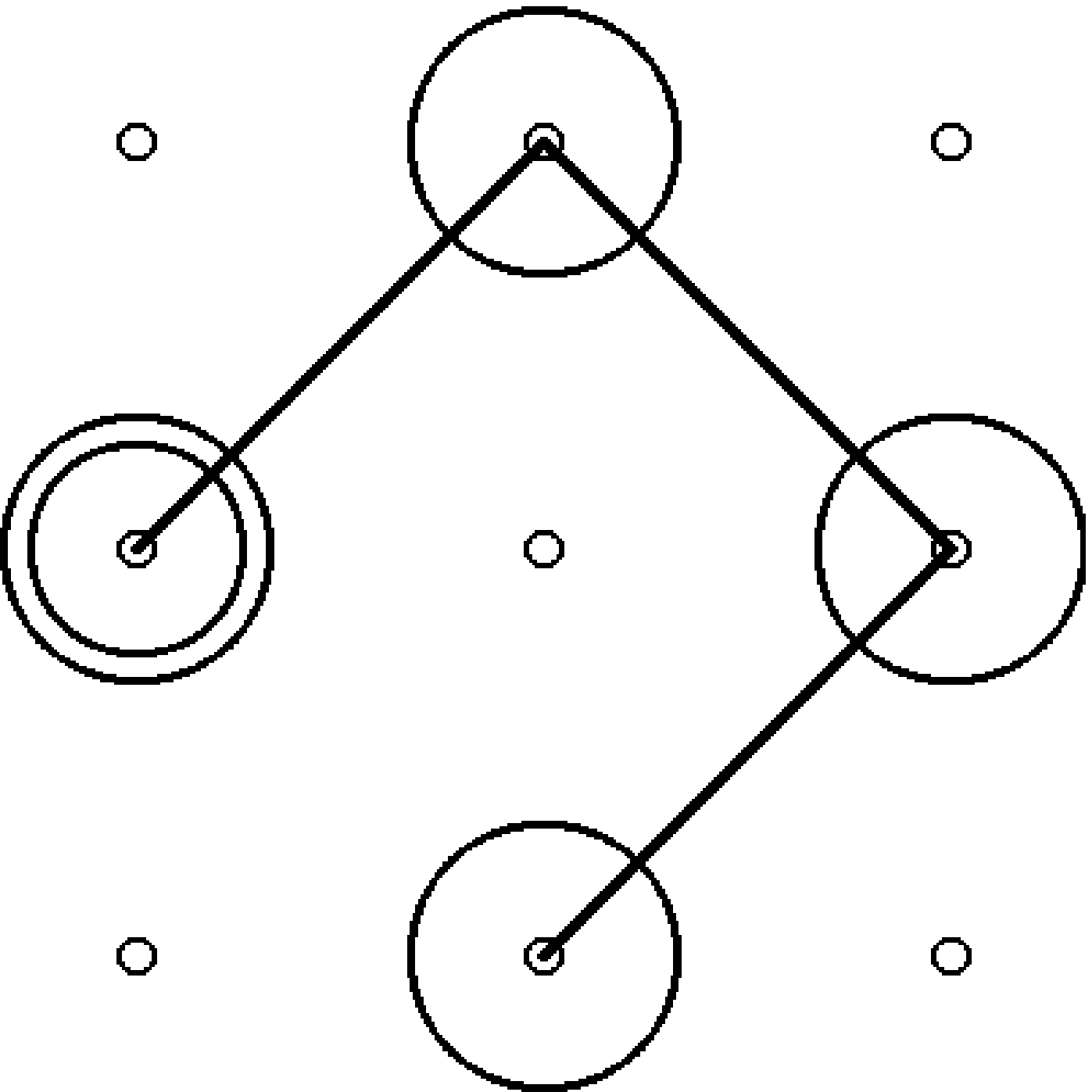}} &
\fbox{\includegraphics[width=0.15\linewidth]{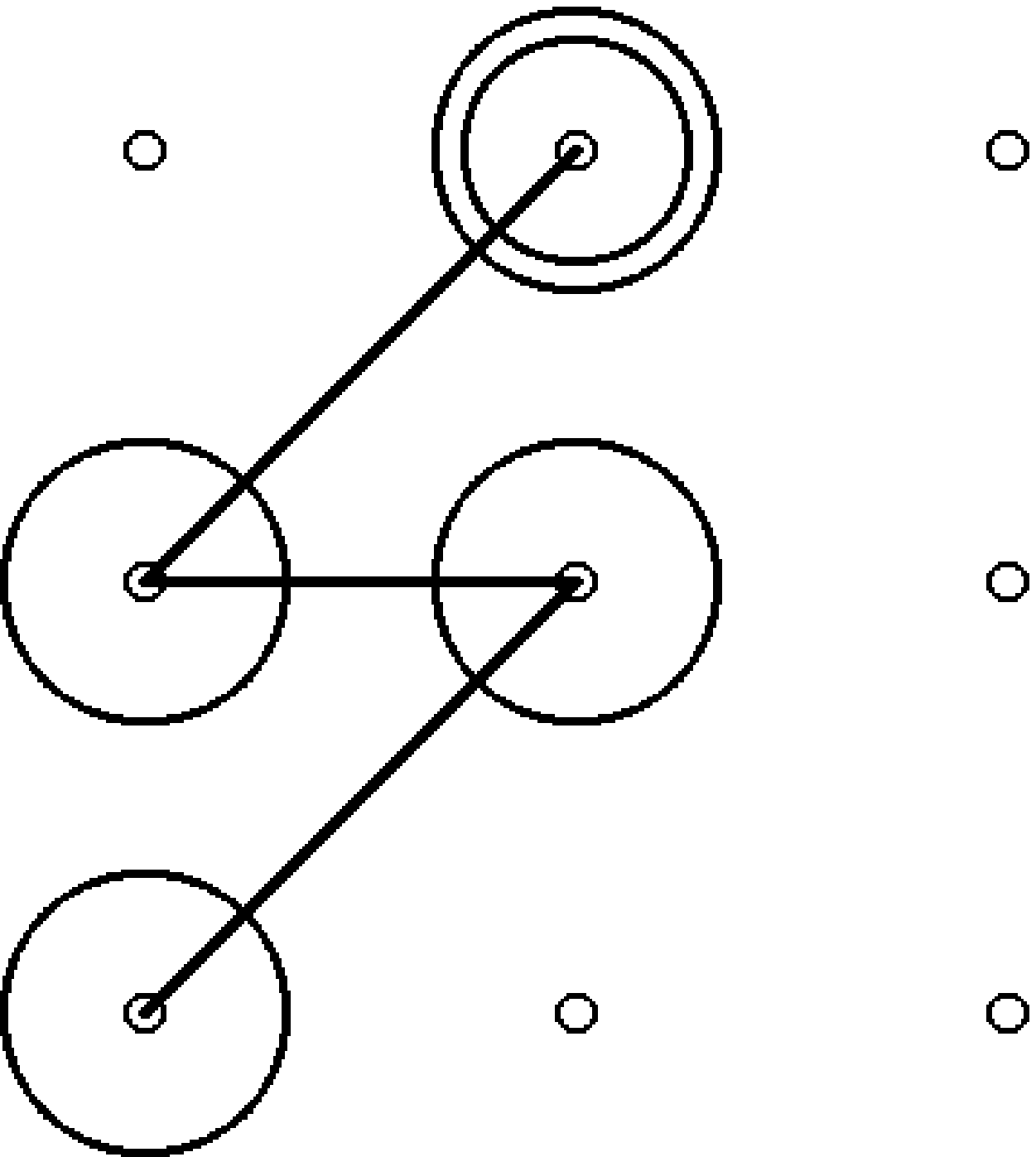}} &
\fbox{\includegraphics[width=0.15\linewidth]{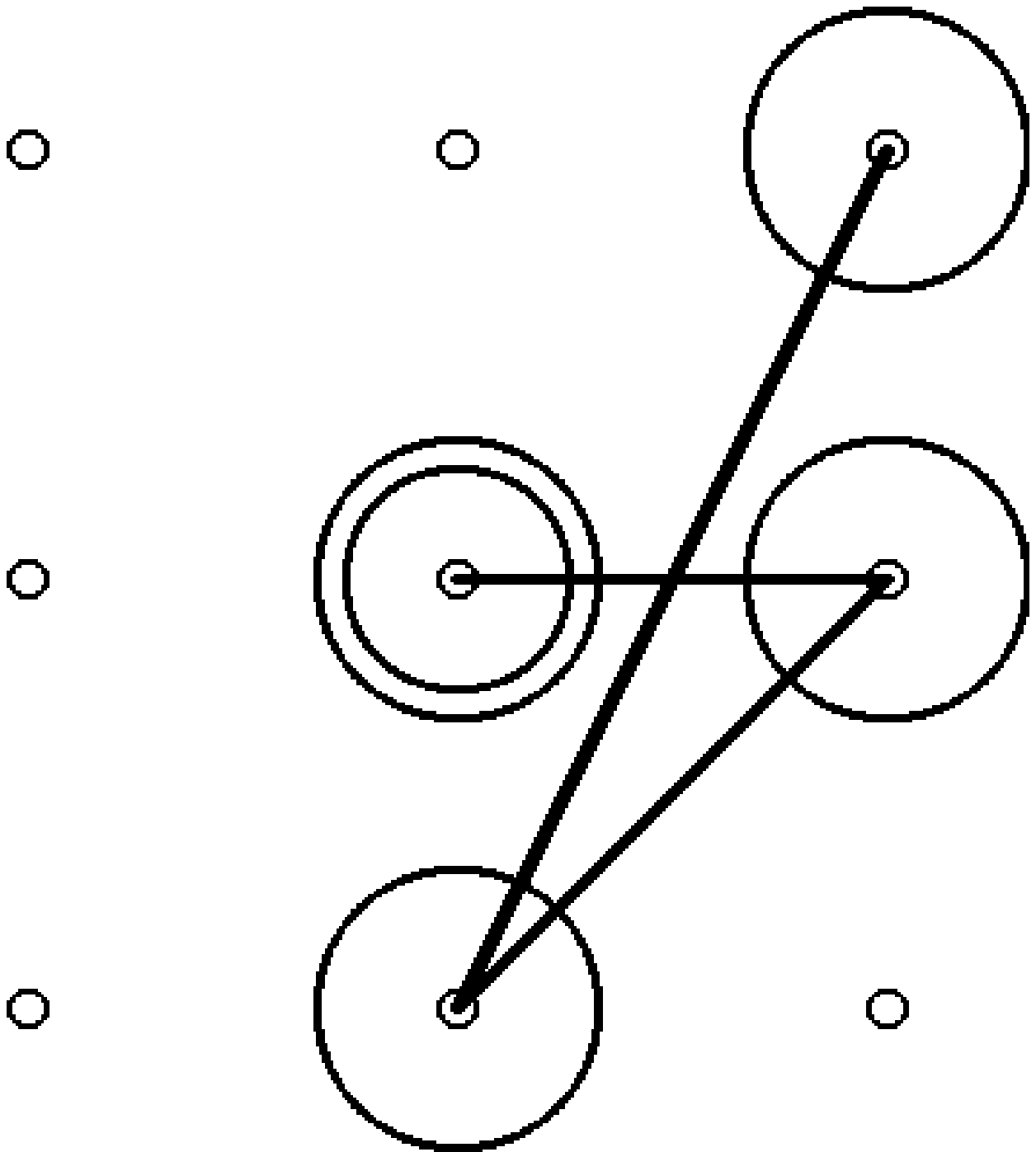}} \\ 
0145 & 6745 & 3157 &  1346 & 4572 \\
\fbox{\includegraphics[width=0.15\linewidth]{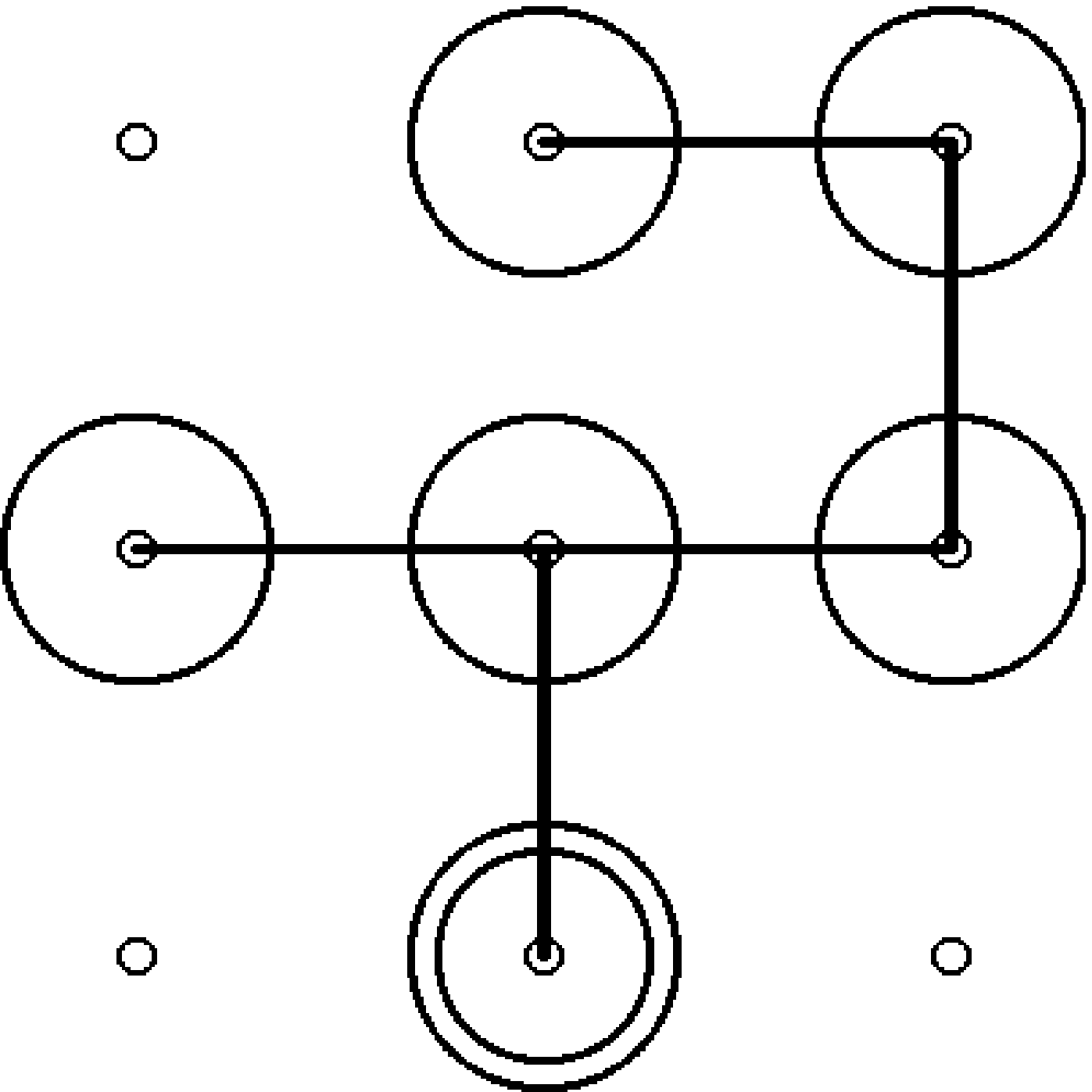}} &
\fbox{\includegraphics[width=0.15\linewidth]{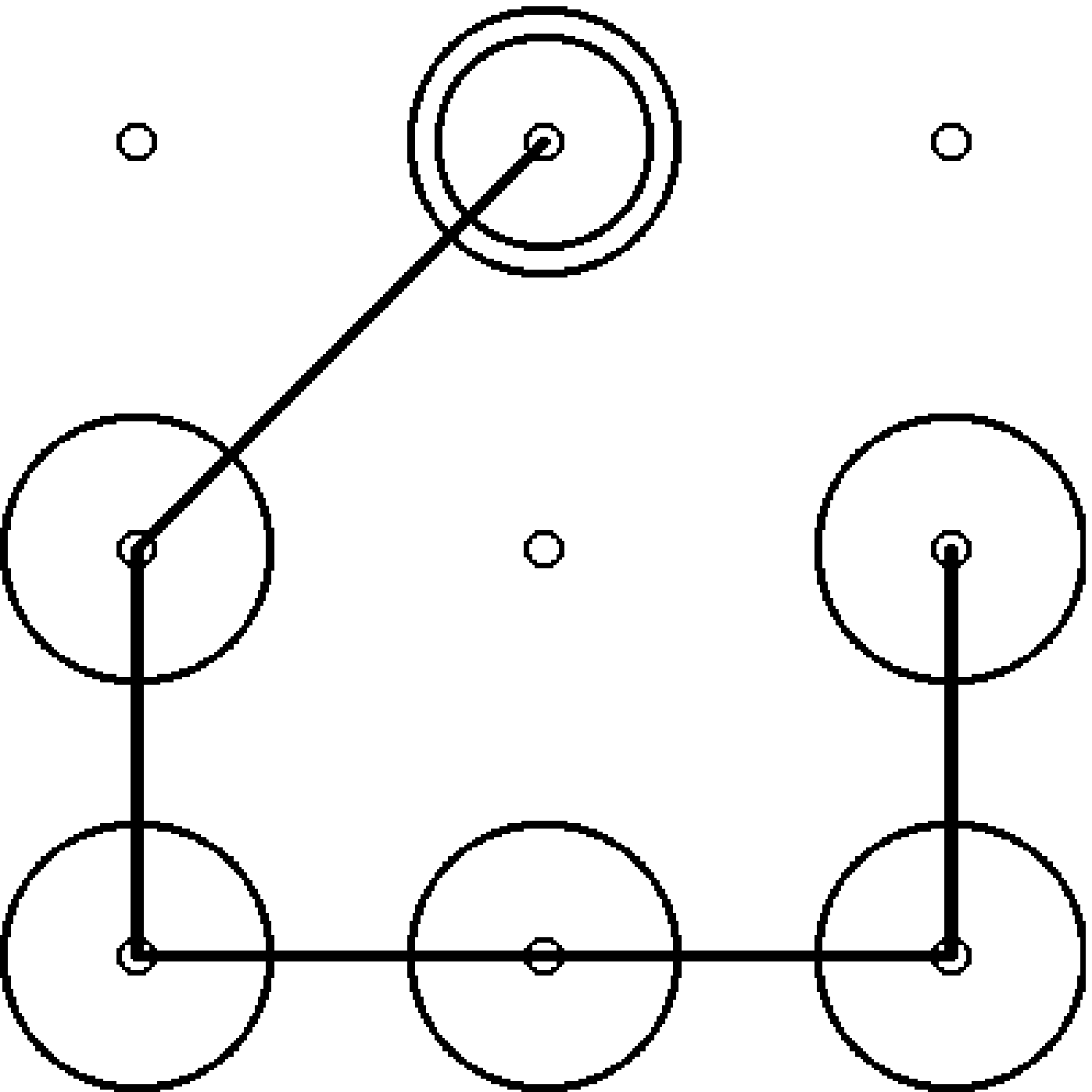}} &
\fbox{\includegraphics[width=0.15\linewidth]{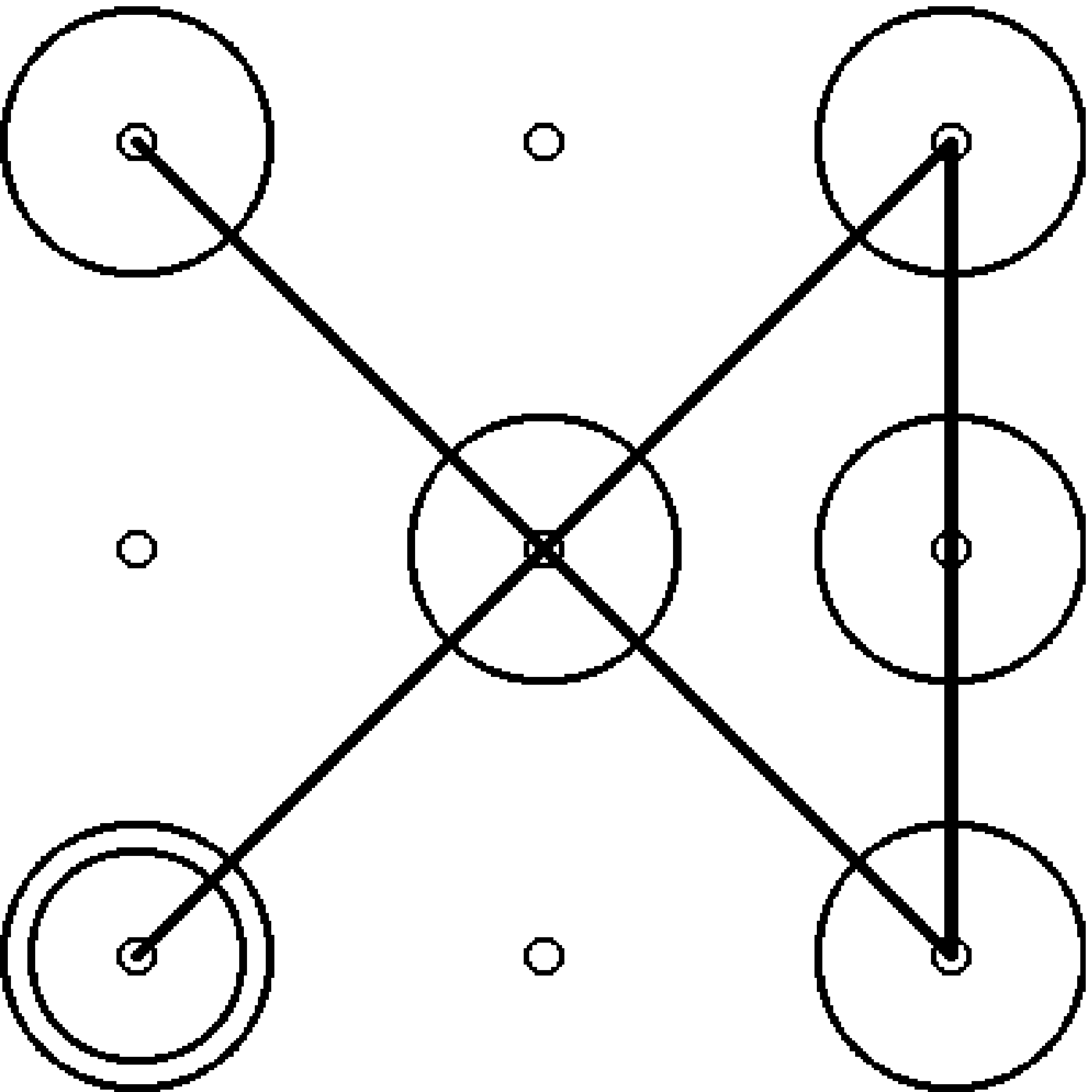}} &
\fbox{\includegraphics[width=0.15\linewidth]{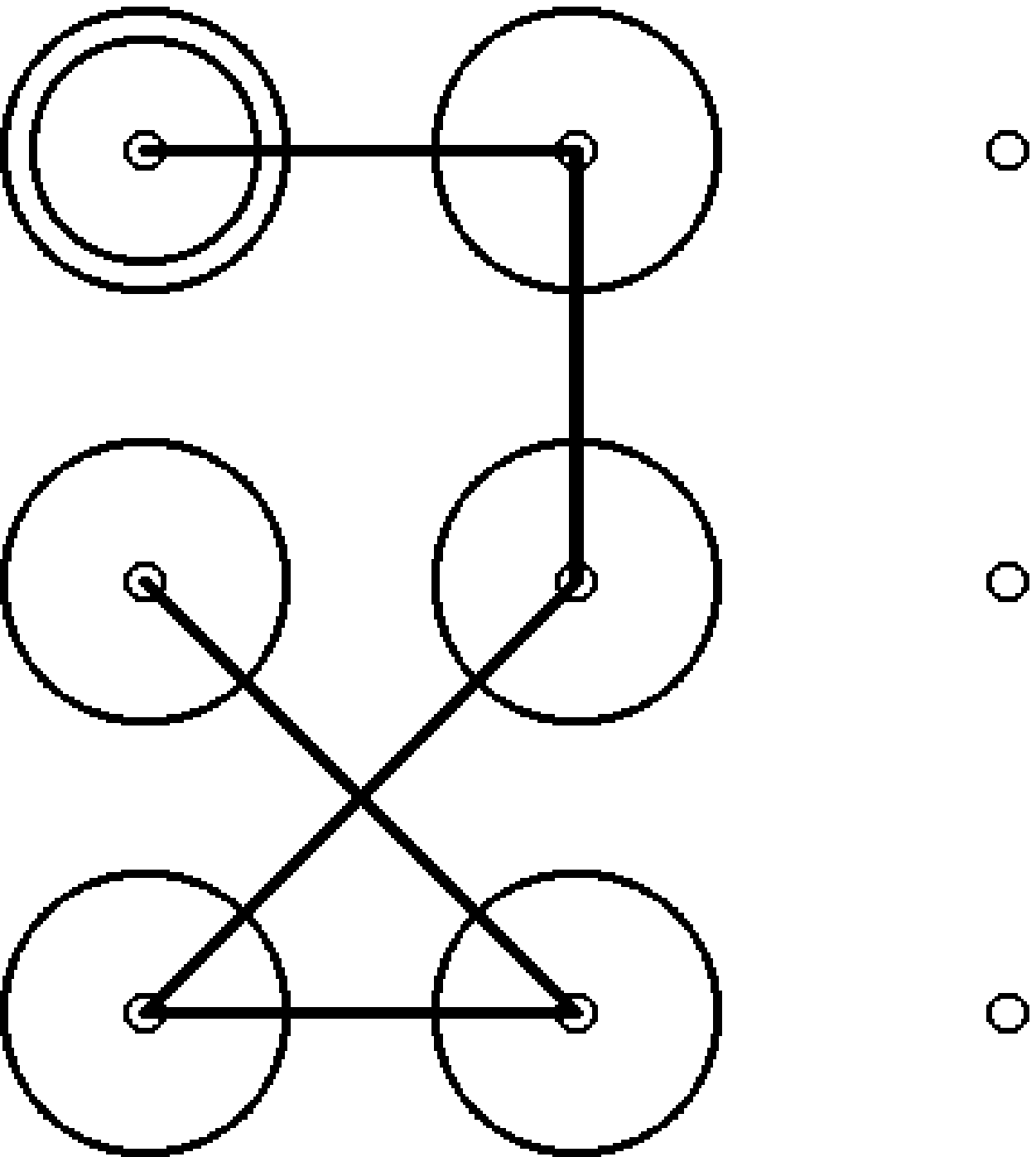}} &
\fbox{\includegraphics[width=0.15\linewidth]{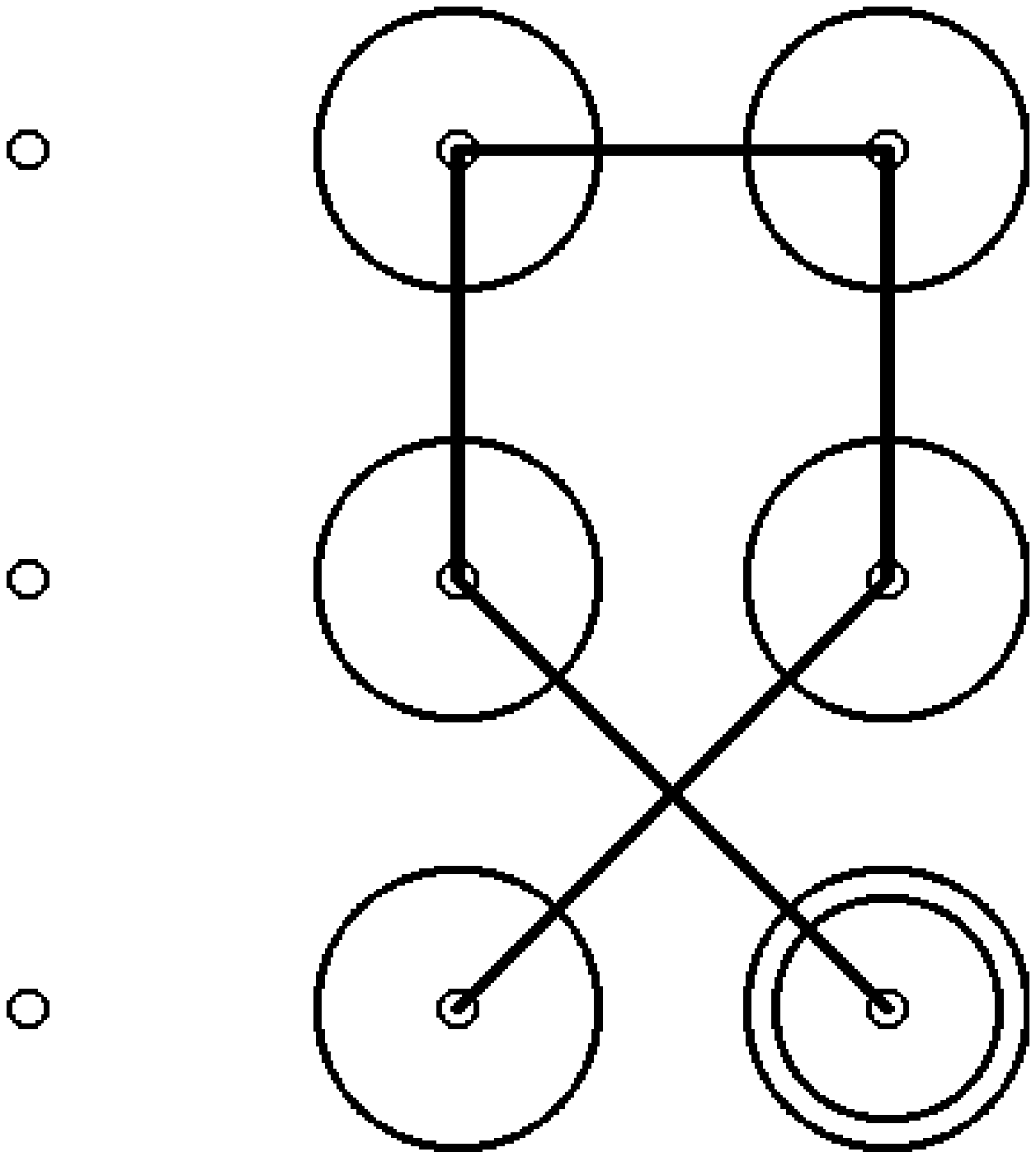}} \\
743521 & 136785 &  642580 & 014673  & 841257 \\
\end{tabular}
\end{center}

\subsection{PINs}
\label{fig:pins}
The filled circle $\bullet$ indicates the start point, and unfilled
circle $\circ$ indicates an intermediate point. Line traces are
provided to show expected shape and directionality of a trace, but
users do not drag/maintain contact during entry. Rather, users enter the PIN as
normally would be expected by clicking/pressing the buttons. All visuals are also
provided in {\tt images/pins} sub-directory.
\begin{center}
  \begin{tabular}{c c c c c}\small
    \fbox{\includegraphics[width=0.15\linewidth]{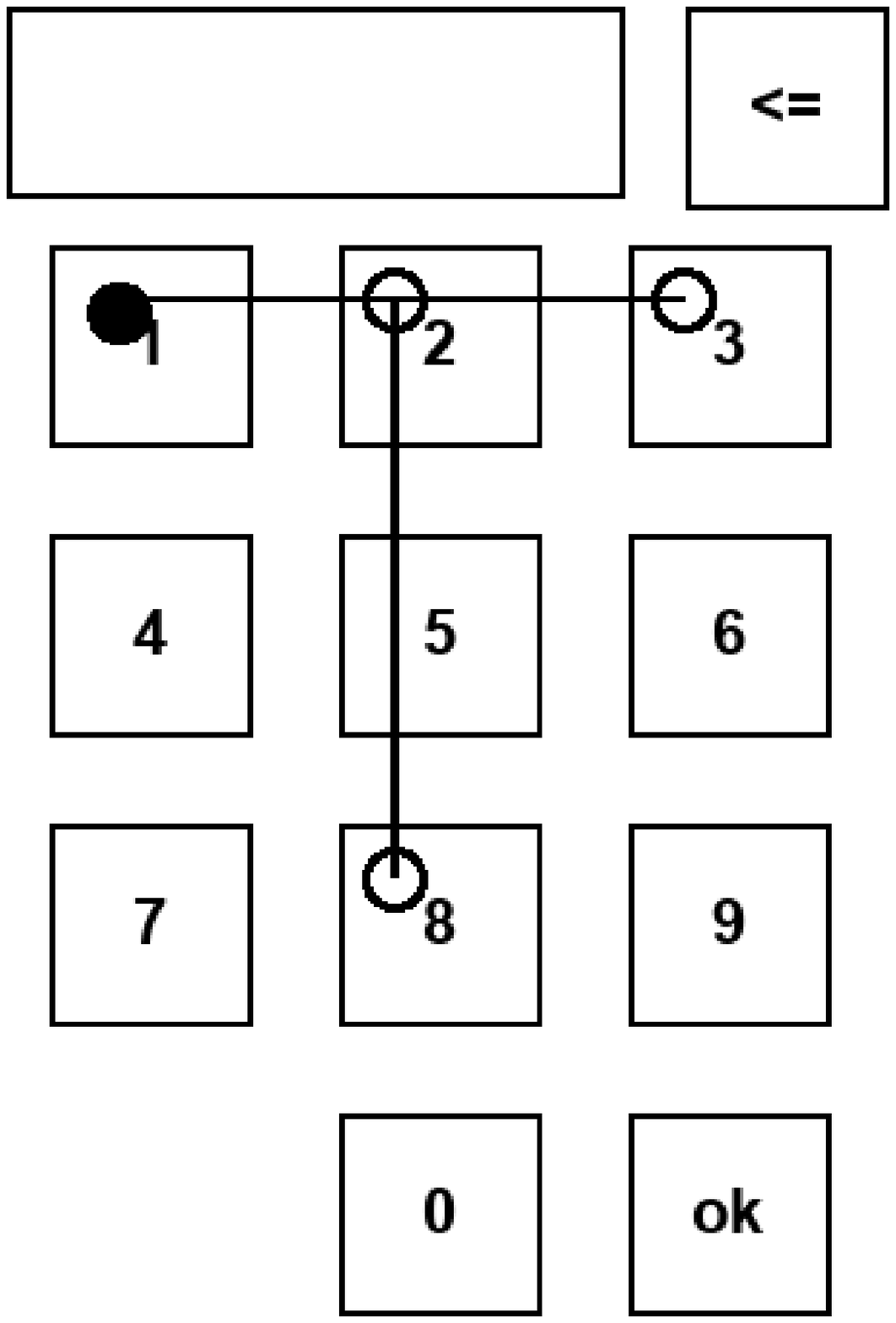}}&
    \fbox{\includegraphics[width=0.15\linewidth]{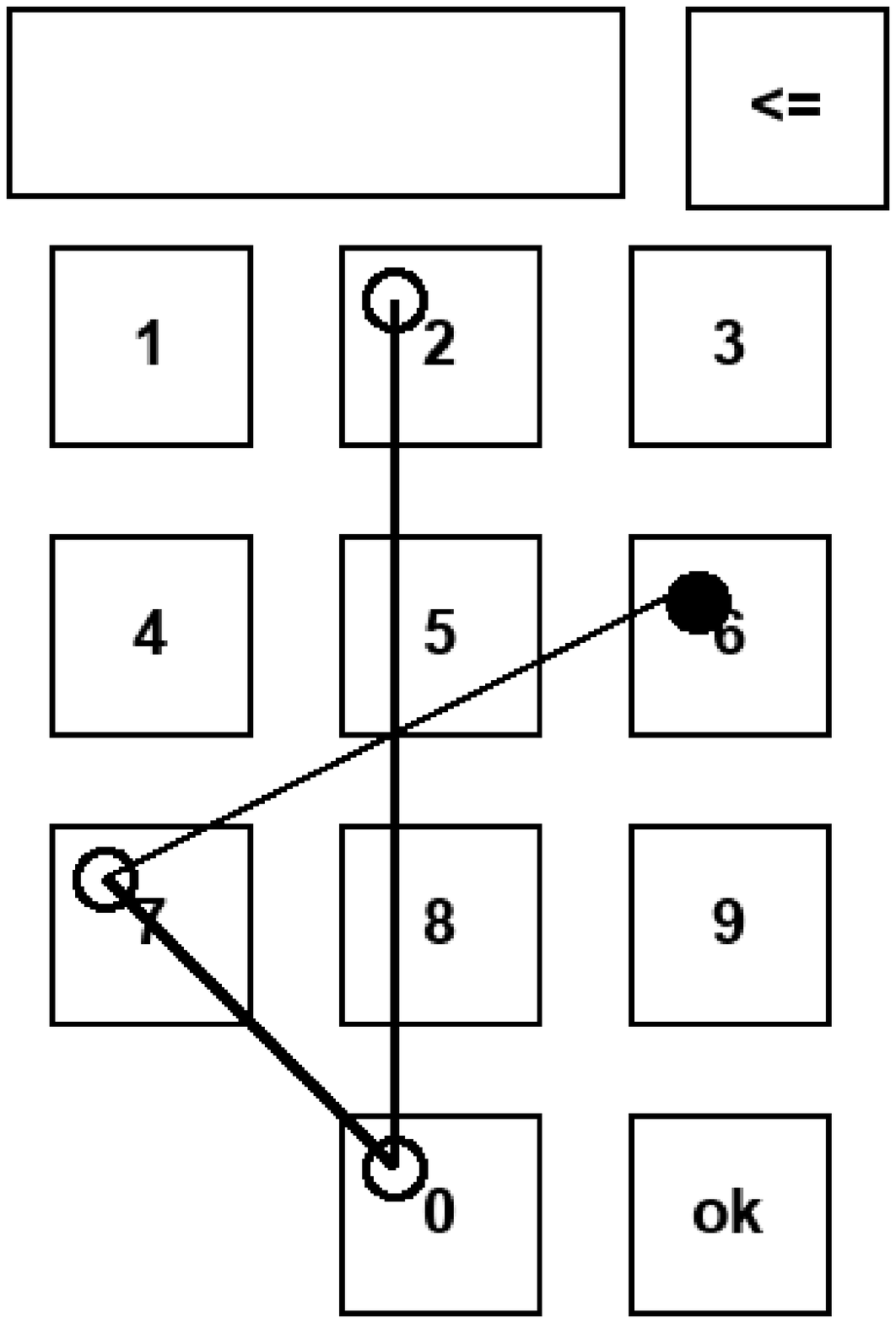}}&
    \fbox{\includegraphics[width=0.15\linewidth]{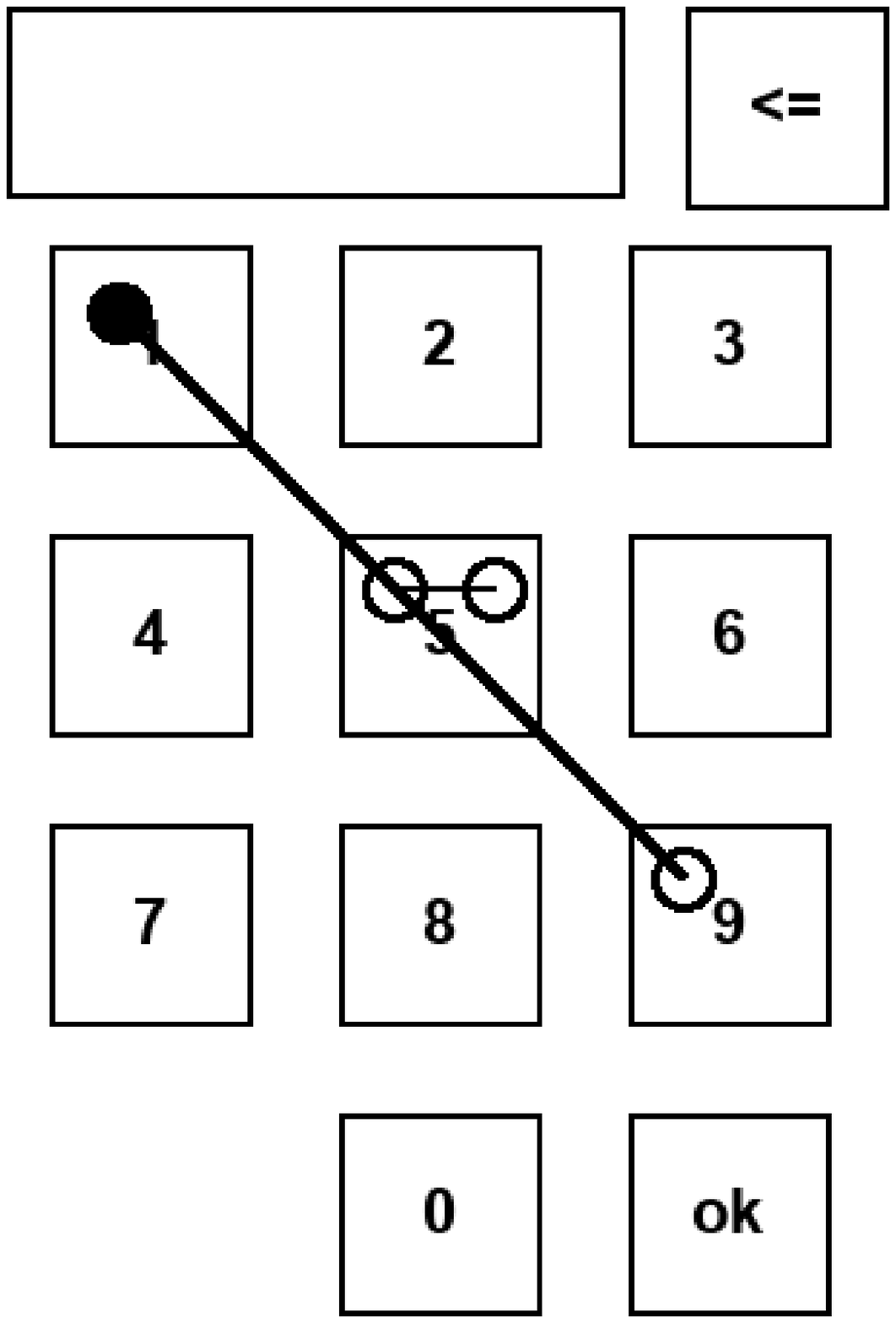}}&
    \fbox{\includegraphics[width=0.15\linewidth]{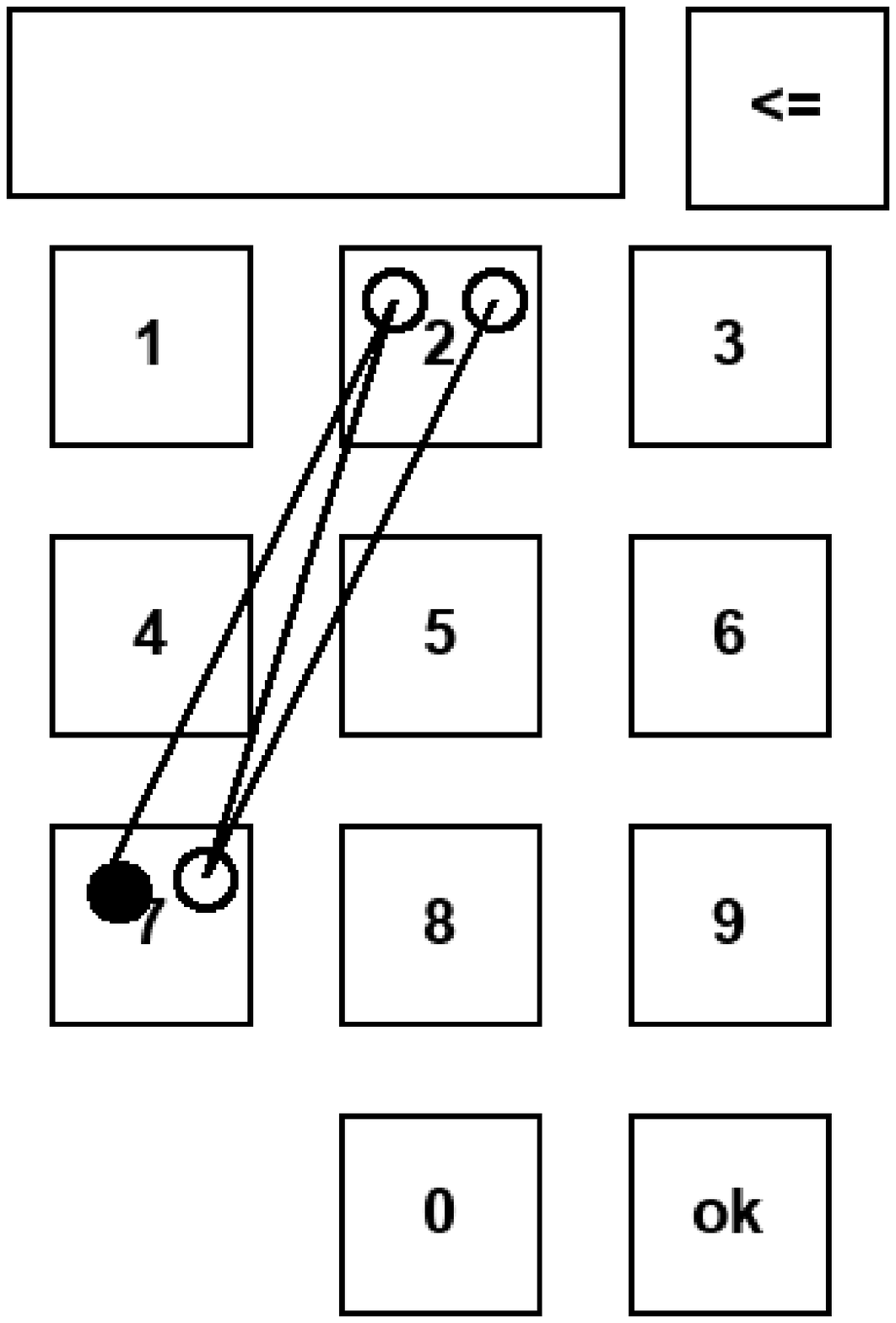}}&
    \fbox{\includegraphics[width=0.15\linewidth]{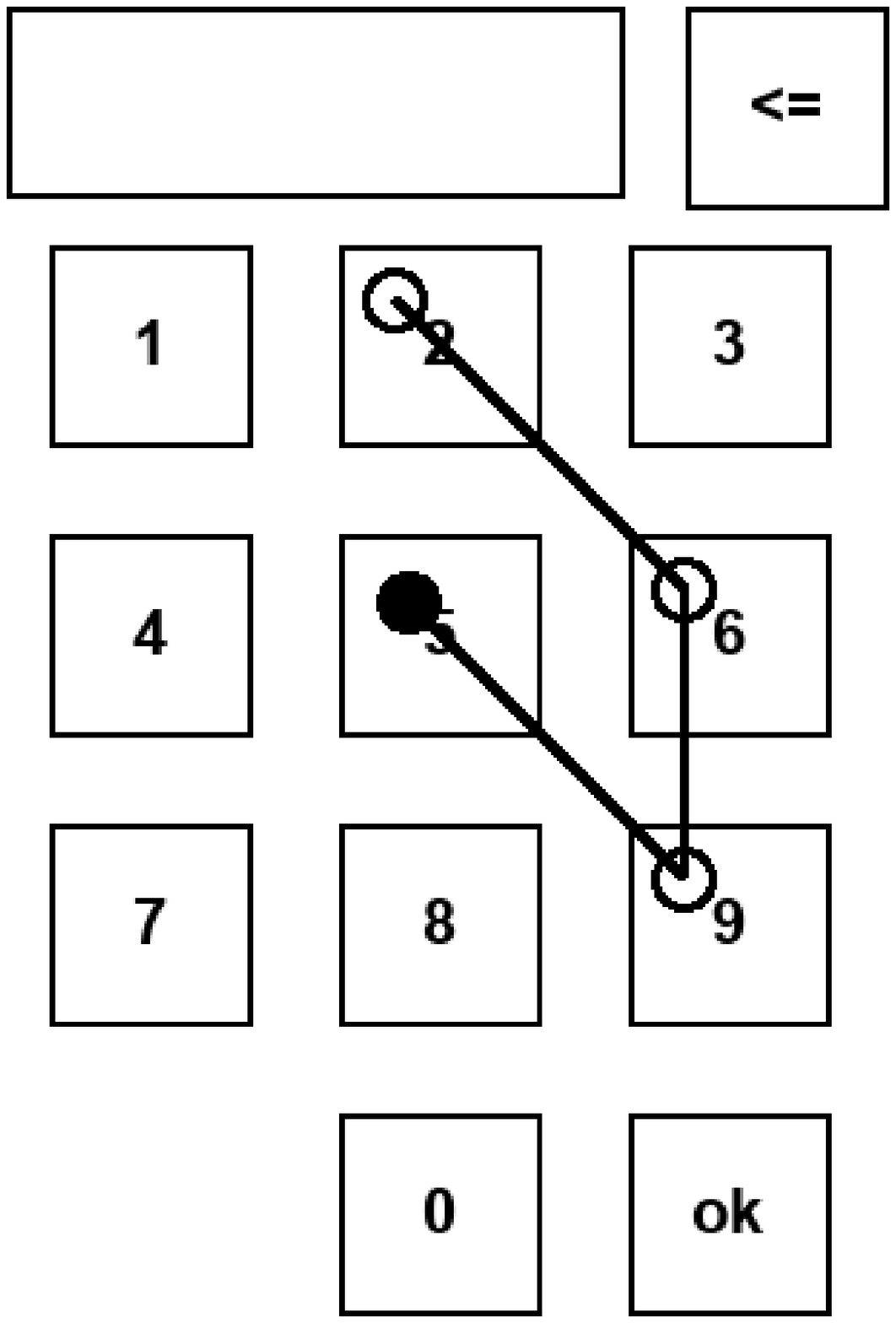}}\\
    1328 & 6702 & 1955    & 7272 & 5962\\
    \fbox{\includegraphics[width=0.15\linewidth]{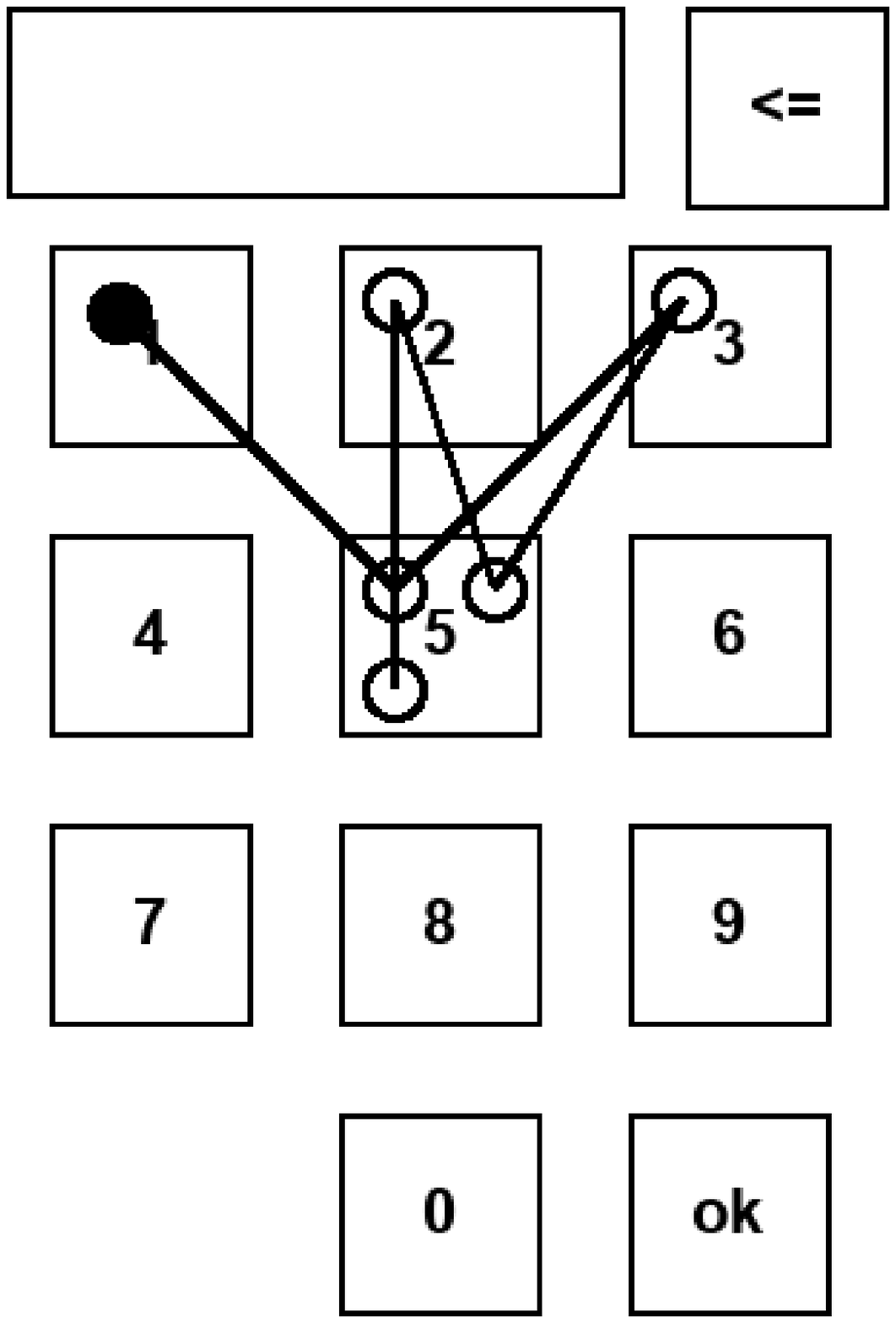}}&
    \fbox{\includegraphics[width=0.15\linewidth]{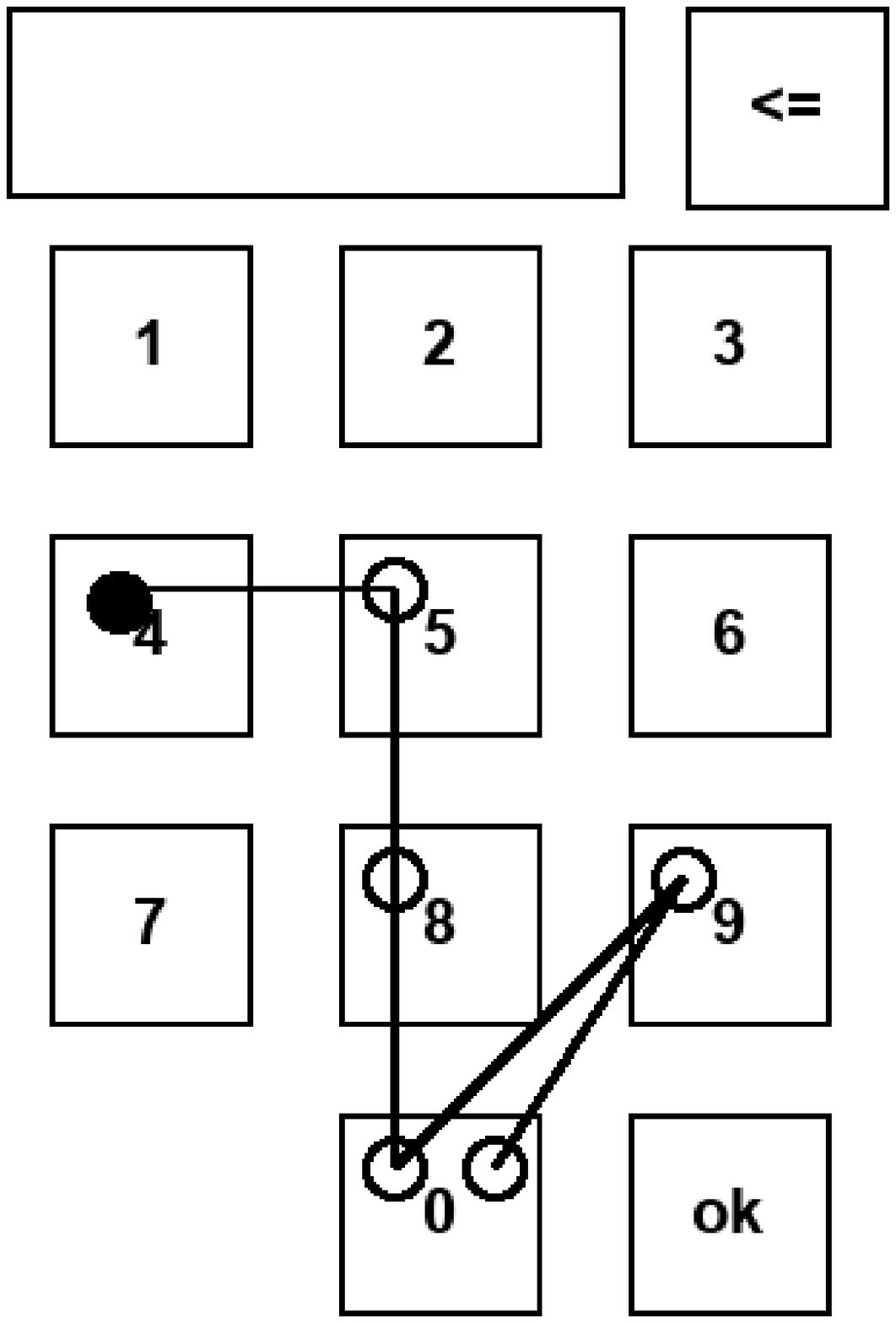}}&
    \fbox{\includegraphics[width=0.15\linewidth]{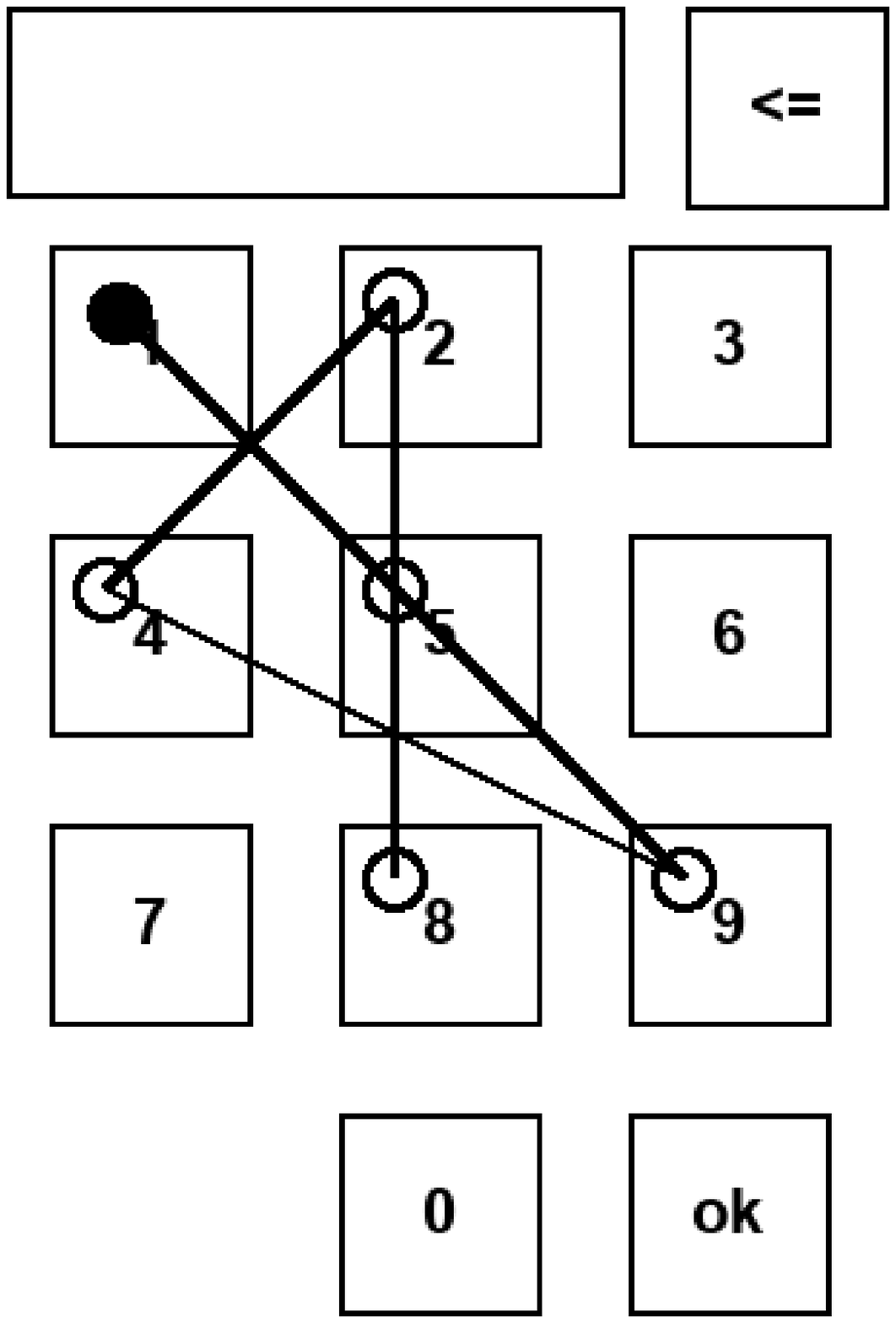}}&
    \fbox{\includegraphics[width=0.15\linewidth]{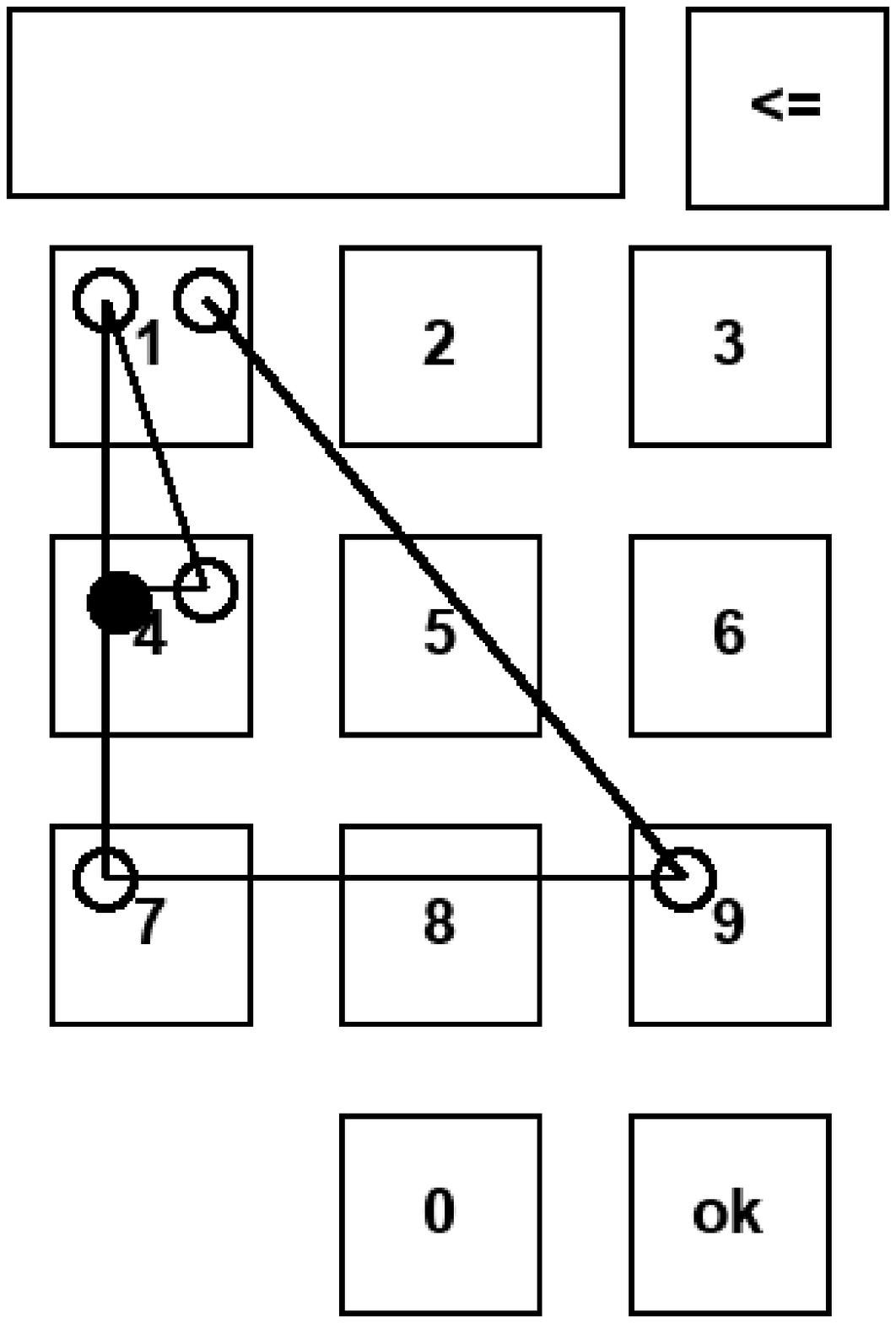}}&
    \fbox{\includegraphics[width=0.15\linewidth]{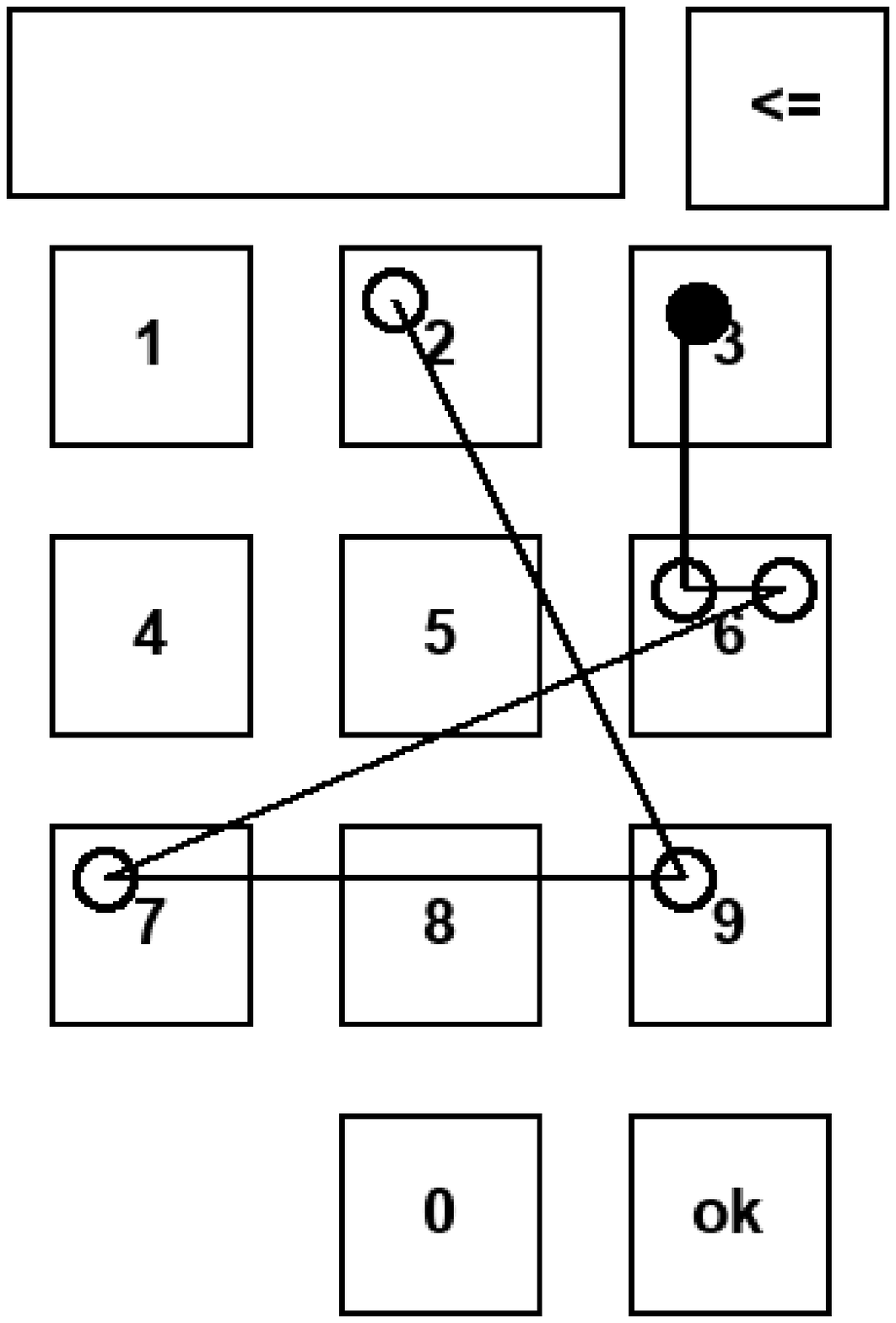}}\\
    152525 & 458090 & 159428 & 441791 & 366792 \\

  \end{tabular}
\end{center}

\section{Pre-Survey Questions}

\begin{itemize}
\item What is your age (18-24, 25-34, 35-44, 45-54, 55+, NA)?

\item What is your identified gender?

\item Do you have any physical conditions that might affect your ability to
  enter authentication passcodes on a mobile phone?

\item Do you use a smartphone currently? What is its operating system? Why did
  you select that phone?

\item Do you use an authentication method to lock your phone, and if so which
  method, and why (i.e. PIN, grid, TouchID, etc.)?

\item Without telling me your current passcode, how do you select the passcodes
  you use to lock your phone (i.e. familiar number, or visual pattern)?

\item How concerned are you with keeping your phone secure (1, not at all
  concerned, to 5, highly concerned)?  
  
  \item What experiences can you recall involving people either trying to steal or use your phone without permission?

\item What experiences can you recall involving people trying to observe your
  passcodes without permission?

\item How concerned are you, typically, in a public space, with the threat of
  someone watching you authenticate and collecting your passcodes (1, not at all
  concerned, to 5, highly concerned)?

\item If you had any of these experiences, how did it affect your behavior?

\item Have any other experiences or concerns indirectly affected your
  authentication behavior (news articles, stories about friends, etc.)?

\item If you do authentication, how do you typically hold your phone for that?

\end{itemize}

\subsection{Post-Survey Questions}
\begin{itemize}
\item On a scale from 1-5, how difficult was entering passcode this way (1, very
  easy, to 5, very hard)? How so?

\item On a scale from 1-5, how easy was the grid pattern tactile app to learn
  (1, very easy, to 5, very hard)? How so?

\item On a scale from 1-5, how easy was the grid pattern tactile app to use (1,
  very easy, to 5, very hard)? How so?

\item On a scale from 1-5, how easy was the PIN tactile app to learn (1, very
  easy, to 5, very hard)? How so?

\item On a scale from 1-5, how easy was the PIN tactile app to use (1, very
  easy, to 5, very hard)? How so?

\item Can you see yourself using the grid pattern tactile aid to help
  authenticate on your phone in your actual daily life? Why or why not?

\item Can you see yourself using the PIN tactile aid to help authenticate on
  your phone in your actual daily life? Why or why not?

\item How is this approach similar or different from how you enter passcodes on
  your phone now?

\item Do you think the grid tactile aid would help protect you from someone
  shoulder surfing you? Why or why not?

\item Do you think the PIN tactile aid would help protect you from someone
  shoulder surfing you? Why or why not?

\end{itemize}



\end{document}